\documentclass[12pt]{article}

\usepackage{amssymb, amsthm, amsmath, amsfonts, eurosym, newtxmath, nccmath}

\usepackage{graphicx}

\usepackage{booktabs, dcolumn, float, adjustbox, multirow}

\usepackage{hyperref}
\usepackage[font = small]{caption}
\usepackage{subcaption}

\usepackage[textsize = small]{todonotes}

\usepackage{algorithm, algpseudocode}
\algrenewcommand\algorithmiccomment[1]{\hfill$\triangleright$~#1} % Right-aligned comments.

\usepackage[backend = biber, style = apa, uniquename = false, sorting = ynt]{biblatex}

\usepackage{geometry, color, setspace, pdflscape, indentfirst}
\usepackage[bottom]{footmisc}

\newcolumntype{L}[1]{>{\raggedright\let\newline\\arraybackslash\hspace{0pt}}m{#1}}
\newcolumntype{C}[1]{>{\centering\let\newline\\arraybackslash\hspace{0pt}}m{#1}}
\newcolumntype{R}[1]{>{\raggedleft\let\newline\\arraybackslash\hspace{0pt}}m{#1}}

\definecolor{linkcolour}{rgb}{0, 0.2, 0.6}
\hypersetup{colorlinks, breaklinks, urlcolor = linkcolour, linkcolor = linkcolour, citecolor = linkcolour}

\newtheorem{assumption}{Assumption}
\newtheorem{theorem}{Theorem}

\usepackage[title]{appendix}

\usepackage{svg}

\makeatletter
\let\save@measuring@true\measuring@true
\def\measuring@true{%
  \save@measuring@true
  \def\beamer@sortzero##1{\beamer@ifnextcharospec{\beamer@sortzeroread{##1}}{}}%
  \def\beamer@sortzeroread##1<##2>{}%
  \def\beamer@finalnospec{}%
}
\makeatother

\DeclareMathOperator{\EX}{\mathbb{E}} % Expected value.
\DeclareMathOperator{\PX}{\mathbb{P}} % Probability.
\newcommand{\indep}{\perp \!\!\! \perp} % Independence.

\newcommand{\covariateSpace}[0]{{\cal X}} 
\newcommand{\potentialOutcome}[1]{Y_{i} ( #1 )} 
\newcommand{\estimatedPotentialOutcome}[1]{\hat{Y}_{i} ( #1 )} 
\newcommand{\conditionalMean}[2]{\mu ( #1, #2 )}
 
\newcommand{\propensityScore}[1]{\pi ( #1 )} 
\newcommand{\estimatedPropensityScore}[1]{\hat{\pi} ( #1 )} 
\newcommand{\normalizedPropensityScore}[1]{\tilde{\pi} ( #1 )} 
\newcommand{\individualEffect}{\xi_i} 
\newcommand{\cate}[1]{\tau ( #1 )} 
\newcommand{\estimatedCate}[1]{\hat{\tau} ( #1 )} 
\newcommand{\normalPdf}[1]{\phi ( #1 )}
\newcommand{\mseCriterion}[3]{MSE ( #1, #2, #3 )}
\newcommand{\tree}[0]{\mathcal{T}}
\newcommand{\trainingSample}[0]{\mathcal{S}^{tr}}
\newcommand{\testSample}[0]{\mathcal{S}^{te}}
\newcommand{\honestSample}[0]{\mathcal{S}^{hon}}
\newcommand{\score}[0]{\psi}

\newcommand{\open}{“} 

\renewcommand{\textcolor}[2]{#2}

\DeclareFontFamily{U}{mathx}{}
\DeclareFontShape{U}{mathx}{m}{n}{<-> mathx10}{}
\DeclareSymbolFont{mathx}{U}{mathx}{m}{n}
\DeclareMathAccent{\widehat}{0}{mathx}{"70}
\DeclareMathAccent{\widecheck}{0}{mathx}{"71}

\bibliography{references}

\begin{document}

%%%%%%%%%%%%%%%%%%%%%%%%%%%%%%%%%%%%%%%%
%            TITLE PAGE.               %
%%%%%%%%%%%%%%%%%%%%%%%%%%%%%%%%%%%%%%%%

\begin{titlepage}
    \title{Aggregation Trees}
    
    \author{Riccardo Di Francesco\thanks{\ Department of Economics, University of Southern Denmark, Odense. Electronic correspondence: rdif@sam.sdu.dk.}}
    
    \date{\today}
    
    \maketitle

    \vspace{-20pt}

    % \begin{center}
    %     \href{https://riccardo-df.github.io/assets/papers/Aggregation_Trees.pdf}{Click here for the most recent version.}
    % \end{center}
    
    \begin{abstract}
        \noindent Uncovering the heterogeneous effects of particular policies or \open treatments" is a key concern for researchers and policymakers. A common approach is to report average treatment effects across subgroups based on observable covariates. However, the choice of subgroups is crucial as it poses the risk of $p$-hacking and requires balancing interpretability with granularity. This paper proposes a nonparametric approach to construct heterogeneous subgroups. The approach enables a flexible exploration of the trade-off between interpretability and the discovery of more granular heterogeneity by constructing a sequence of nested groupings, each with an optimality property. By integrating our approach with \open honesty" and debiased machine learning, we provide valid inference about the average treatment effect of each group. We validate the proposed methodology through an empirical Monte-Carlo study and apply it to revisit the impact of maternal smoking on birth weight, revealing systematic heterogeneity driven by parental and birth-related characteristics.
                
        \vspace{6pt}
        

        \bigskip
    \end{abstract}
    
    \setcounter{page}{0}
    
    \thispagestyle{empty}
\end{titlepage}

\pagebreak \newpage

\doublespacing

%%%%%%%%%%%%%%%%%%%%%%%%%%%%%%%%%%%%%%%%
%            INTRODUCTION.             %
%%%%%%%%%%%%%%%%%%%%%%%%%%%%%%%%%%%%%%%%

\section{Introduction}
\label{sec_introduction}
\noindent Understanding the effects of a particular policy or \open treatment" is a key concern for researchers and policymakers. Traditionally, the assessment of the policy's actual effectiveness involves the identification and estimation of the Average Treatment Effect (ATE), a parameter that quantifies the average impact of the policy on the reference population \parencite[see, e.g.,][]{angrist2009mostly, imbens2015causal}. However, while the ATE is straightforward to interpret, it lacks information regarding effect heterogeneity and therefore does not allow us to explore the distributional impacts of the policy, which hold significant importance for decision-making when the social welfare criterion representing the preferences of the policymakers is not \open utilitarian" \parencite[][]{kitagawa2021equality}.

A common approach to tackle effect heterogeneity is to report the ATEs across different subgroups defined by observable covariates. These Group Average Treatment Effects (GATEs) enable us to explore heterogeneity while maintaining a certain level of interpretability and are widely employed in applied research.\footnote{\ For instance, \textcite{chernozhukov2017generic} document that, among 189 randomized control trials published in top economic journals since 2006, $40\%$ report at least one subgroup analysis.} GATEs are particularly valuable when decision rules need to be applied or interpreted by humans, such as treatment guidelines for physicians \parencite[see, e.g.,][]{athey2016recursive}. \textcolor{purple}{However, two practical issues arise. First, there could be many ways to form subgroups, and when no natural choice exists, iteratively searching for subgroups with significantly estimated GATEs raises the possibility of $p$-hacking \parencite[see, e.g.,][]{imbens2021statistical}.\footnote{\ \textcolor{purple}{\ In some empirical applications, subgroup definitions are \open natural" because they are guided by domain knowledge or policy relevance. Examples include clinically meaningful age bands, and gender or ethnicity strata.}} Second, even when subgroup definitions are fixed, deciding how many groups to report remains nontrivial: more groups can reveal finer heterogeneity, but too many can undermine interpretability.}

This paper introduces a methodology for constructing heterogeneous subgroups that enables a flexible and coherent exploration of the trade-off between interpretability and the discovery of more granular heterogeneity while avoiding complications associated with $p$-hacking. \textcolor{purple}{The approach can serve as an alternative to pre-analysis plans---often criticized for limiting the potential for uncovering unexpected heterogeneity---by enabling agnostic exploration of effect heterogeneity. It can also complement pre-specified analyses by providing a data-driven partition to assess subgroup patterns beyond the pre-registered hypotheses.}

The proposed methodology, hereafter referred to as \textit{aggregation trees}, builds on standard decision trees \parencite[][]{breiman1984cart} to aggregate units with similar estimated responses to the treatment.\footnote{\ \textcolor{purple}{Essentially, we adopt a “fit-the-fit” strategy \parencite[see, e.g.,][]{hahn2020bayesian, bargagli2020causal, bargagli2022heterogeneous}: first estimate CATEs using any suitable method, then fit these estimates with a decision tree.}} The resulting tree is then pruned to generate a sequence of groupings, one for each level of granularity. We show that each grouping features an optimality property in that it ensures that the loss in explained heterogeneity resulting from aggregation is minimized. Moreover, the sequence is nested in the sense that subgroups formed at a particular level of granularity are never disrupted at coarser levels. This property guarantees the consistency of the results across the different granularity levels, which is a fundamental requirement for any classification system \parencite[see, e.g.,][]{cotterman1992classification}.

For a particular grouping, \textcolor{purple}{we leverage debiased machine learning procedures \parencite{semenova2021debiased} to obtain point estimates and standard errors for the GATEs.\footnote{\ \textcolor{purple}{In randomized experiments, a simple regression of outcomes on group dummies and their interactions with treatment assignment ensures that the interaction coefficients identify each group’s GATE.}} We further combine this approach with \open honesty" \parencite[][]{athey2016recursive} to deliver valid inference. Honesty is a subsample-splitting technique that requires that different observations are used to form subgroups and estimate the GATEs. In analogy to classical econometrics, this is equivalent to using different subsamples to select and estimate a model. This way, the asymptotic properties of GATE estimates are the same as if the groupings had been exogenously given, and we can use the estimated standard errors to conduct valid inference as usual, e.g., by constructing conventional confidence intervals.}

\textcolor{purple}{Specifically, we use a \textit{training subsample} to estimate CATEs with any suitable method and to construct the sequence of groupings. We then use a disjoint \textit{honest subsample} to estimate the GATEs. To account for selection into treatment, we construct a standard Neyman-orthogonal score using cross-fitted nuisance functions \parencite{chernozhukov2018double} and regress this score on group dummies, performing all steps on the honest sample to maintain the inferential guarantees of \textcite{semenova2021debiased}.\footnote{\ \textcolor{purple}{Neyman-orthogonal scores are central in recent causal machine learning literature. \textcite{chernozhukov2018double} show their advantages for ATE estimation and inference with flexible machine learning estimation of the nuisance functions. \textcite{semenova2021debiased} extend this logic to provide estimation and inference methods for the best linear predictor of the CATE function, which automatically target GATEs when the chosen set of basis function consists of group dummies. \textcite{kennedy2023towards} pushes this further by combining orthogonal scores with linear smoothing techniques---shown to satisfy a key \open stability" condition---to construct a two-stage doubly robust CATE estimator.}}}

Our methodology is similar to the causal trees of \textcite{athey2016recursive}, but it differs in its direct applicability to observational studies and its focus on the trade-off between interpretability and the discovery of more granular heterogeneity. We compare the performance of aggregation and causal trees using an empirical Monte-Carlo study \parencite[][]{huber2013performance, lechner2013sensitivity}.\footnote{\ The literature presents a wide array of causal machine learning methodologies for estimating dense heterogeneous treatment effects \parencite[see, e.g.,][]{wager2018estimation, athey2019generalized, kunzel2019metalearners, lechner2022modified}. However, aggregation trees focus on the construction of heterogeneous subgroups and the subsequent estimation of GATEs. Given these distinct objectives, we do not include these methodologies in our simulations.} Our simulation shows that aggregation trees lead to lower root mean squared error in the estimated treatment effects, with reductions of up to $121\%$. This improvement entirely stems from the lower variance of aggregation trees, resulting from a splitting strategy that is robust to covariates affecting the outcome levels but not the treatment effects.

We also investigate the benefits of honesty compared to more standard \open adaptive" estimation that uses the same data for constructing the tree and GATE estimation. Honesty greatly benefits inference, ensuring approximately nominal coverage of confidence intervals. In contrast, adaptive estimation can result in coverage rates as low as $58\%$. 

The proposed methodology is applied to revisit the impact of maternal smoking on birth weight \parencite[see, e.g.,][]{almond2005costs, cattaneo2010efficient}. The analysis finds evidence of systematic heterogeneity, as different subgroups react differently to the same treatment. Moreover, the analysis reveals that effect heterogeneity is driven by parental and birth-related characteristics. The results are consistent with previous research showing that the effects are stronger for children born to adult mothers \parencite[][]{abrevaya2015estimating, zimmert2019nonparametric}. Furthermore, we provide evidence that the effects are more pronounced when prenatal care visits are more frequent and occur earlier.

\textcolor{purple}{This paper contributes to three distinct strands of the literature. First, it relates to tree-based subgroup-discovery methodologies. These approaches use recursive partitioning of the covariate space to construct heterogenous groups, adapting the standard CART algorithm \parencite{breiman1984cart} to target treatment effects rather than outcomes. For instance, \textcite{athey2016recursive} grow trees by choosing splits that minimize an estimate of the mean-squared error of treatment effects and employ sample-splitting techniques---i.e., honesty---for valid inference, while \textcite{steingrimsson2019subgroup} maximize standardized differences in treatment effects using covariate-adjusted leaf estimators.\footnote{\ \textcolor{purple}{These approaches are developed for randomized experiments. Extensions to selection-on-observables \parencite{yang2022causal} and to instrumental-variables settings \parencite{bargagli2020causalIV} exist.}} A complementary line of research leverages tree-ensemble algorithms \parencite[see, e.g.,][]{breiman2001random, chen2016xgboost} to reduce instability and explore richer partitions.\footnote{\ \textcolor{purple}{For example, \textcite{bargagli2020causal} use ensembles of decision trees to generate a large set of candidate subgroups---defined by if-then decision rules---and then select those most predictive of preliminary CATE estimates using LASSO \parencite{tibshirani1996regression}.}} Yet single-tree models remain more interpretable, facilitating communication with non-experts and direct use in regulatory policy \parencite[see, e.g.,][]{lee2021discovering, bargagli2022heterogeneous} and in learning treatment-assignment policies \parencite{athey2021policy, bodory2024enabling}. This paper contributes by introducing a novel tree-based methodology for constructing heterogeneous subgroups that applies standard CART to estimated CATEs and combines it with honesty \parencite{athey2016recursive} and debiased machine learning \parencite{semenova2021debiased} to deliver valid inference for GATEs.}

\textcolor{purple}{Second, this work complements methods that estimate heterogeneous treatment effects. The recent causal machine learning literature adapts machine learning tools to CATE estimation, most naturally under selection-on-observables with many covariates.\footnote{\ \textcolor{purple}{Broadly, there are two main strategies. One decomposes the problem into supervised prediction tasks via meta-learners \parencite[see, e.g.,][]{kunzel2019metalearners}. The other tailors machine learning algorithms to produce causal estimates directly rather than outcome predictions \parencite{wager2018estimation, athey2019generalized, lechner2018modified, lechner2022modified, hahn2020bayesian}.}} Yet unit-level estimates can be hard to interpret and may exhibit substantial sampling variability, so what appears as heterogeneity may simply be estimation noise \parencite[see, e.g.,][]{chernozhukov2017generic}. This has motivated GATE analyses and associated methods \parencite[see, e.g.,][]{abrevaya2015estimating, lee2017doubly, lechner2018modified, zimmert2019nonparametric, fan2022estimation, lechner2022modified}, which, however, require researchers to predefine groups.\footnote{\ \textcolor{purple}{\textcite{bearth2024causal} discuss how to analyze and interpret differences in GATEs across groups while accounting for variation in other covariates. \textcite{lee2021discovering} develop randomization-based tests to assess whether subgroups defined by a given tree have GATEs that differ from the overall ATE.}} Our contribution is to propose a methodology for constructing subgroups from the data---thus removing the need for ex ante group definitions---and then providing GATE estimation and inference via honesty \parencite{athey2016recursive} and debiased machine learning \parencite{semenova2021debiased}.}

\textcolor{purple}{Third, our empirical findings relate to the broader literature on maternal risky behaviors and birth outcomes. Maternal behaviors are important policy levers because they are modifiable risk factors.\footnote{\ \textcolor{purple}{\textcite{bhalotra2017infant} find that interventions providing information and support to mothers improved both short- and long-run infant survival.}} Because smoking during pregnancy is widely considered the most salient---and most readily modifiable---of these risks \parencite{almond2005costs}, a large literature examines its impact on infant health.\footnote{\ \textcolor{purple}{\textcite{bhalotra2019twin} show that smoking during pregnancy—along with other maternal health conditions and risky behaviors—is negatively associated with the probability of twin birth.}} Numerous studies consistently document sizable negative average effects on birth weight \parencite[see, e.g.,][]{almond2005costs, abrevaya2006estimating} and increasingly negative effects with maternal age \parencite{abrevaya2015estimating, lee2017doubly, zimmert2019nonparametric, fan2022estimation}. Differences in smoking behavior also help explain variation in effects \parencite{cattaneo2010efficient, heiler2021effect, bodory2022high}. This paper contributes by providing robust evidence of systematic heterogeneity---different subgroups of infants are affected differently by maternal smoking---and by offering novel evidence that effects are more pronounced when prenatal care begins earlier.}

The rest of the paper unfolds as follows. Section \ref{sec_framework} discusses the estimands of interest and their identification. Section \ref{sec_aggregation_trees} introduces aggregation trees and compares them to causal trees. Section \ref{sec_empirical_montecarlo} shows the simulation results. Section \ref{sec_empirical_illustration} illustrates the empirical exercise. Section \ref{sec_conclusion} concludes. 

%%%%%%%%%%%%%%%%%%%%%%%%%%%%%%%%%%%%%%%%
%           CAUSAL FRAMEWORK.          %
%%%%%%%%%%%%%%%%%%%%%%%%%%%%%%%%%%%%%%%%

\section{Causal framework} 
\label{sec_framework}

%%% ESTIMANDS.
\subsection{Estimands}
\label{subsec_framework_estimands}
\noindent We define the estimands of interest using the potential outcomes model \parencite{neyman1923, rubin1974estimating}. Suppose to have access to a sample of $n$ i.i.d. observations $( Y_i, D_i, X_i )$, where $Y_i \in \mathcal{Y}$ is the outcome targeted by the treatment, $D_i \in \{ 0, 1 \}$ is the binary treatment indicator, and $X_i = ( X_{i1}, \dots, X_{ip} )^\top \in \covariateSpace$ is a $p \times 1$ vector of pre-treatment covariates. We posit the existence of two potential outcomes $\potentialOutcome{0}$ and $\potentialOutcome{1}$, representing the outcome that the $i$-th unit would experience under each treatment level. The observed outcome for unit $i$ is then the potential outcome corresponding to the treatment received:\footnote{\ The definition of potential outcomes and the observational rule linking them to the observed outcomes implicitly assume the absence of spillover effects among units, which is violated in settings where some units are connected through networks.}
\begin{equation}    
    Y_i = D_i \potentialOutcome{1} + ( 1 - D_i ) \potentialOutcome{0}.
    \label{equation_sutva}
\end{equation}

To define the effect of the treatment, we can take the differences in the potential outcomes of each unit $\individualEffect := \potentialOutcome{1} - \potentialOutcome{0}$ and aggregate them at different levels of granularity. The coarsest estimand of interest is the Average Treatment Effect (ATE), 
\begin{equation}
    \tau := \EX [ \individualEffect ].
    \label{equation_ate}
\end{equation}
\noindent The ATE quantifies the average impact of the policy on the reference population and is straightforward to interpret. However, it lacks information regarding the  distributional impacts of the policy. 

To tackle effect heterogeneity, we can focus instead on the Conditional Average Treatment Effects (CATEs), 
\begin{equation}
    \cate{X_i} := \EX [ \individualEffect | X_i ].
    \label{equation_cates}
\end{equation}
\noindent The CATEs provide information at the finest level of granularity achievable with the information at hand and enable us to relate effect heterogeneity to the observable covariates. However, they are difficult to interpret.

The Group Average Treatment Effects (GATEs) provide a way to explore heterogeneity while maintaining a certain level of interpretability. The GATEs are \textcolor{purple}{averages of (potentially heterogeneous) individual treatment effects within regions of the covariate space and are} defined by
\begin{equation}
    \tau_g := \EX [ \individualEffect | X_i \in \covariateSpace_g ], \quad g = 1, \dots, G,
    \label{equation_gates}
\end{equation}
\noindent where the groups $\covariateSpace_1, \dots, \covariateSpace_G$ represent a partition of $\covariateSpace$.\footnote{\ If grouping is based on the levels of a single discrete variable $Z_i \subset X_i$, each GATE simplifies to $\tau_g = \EX [ \individualEffect | Z_i = g ]$.} \textcolor{purple}{Importantly, equation \eqref{equation_gates} allows for heterogeneous treatment effects within groups: we do not assume $\xi_i = \xi_j$ for units $i$ and $j$ with $X_i, X_j \in \covariateSpace_g$. We also do not posit a \open true" partition of $\covariateSpace$; rather, GATEs are used as an interpretable summary of effect heterogeneity.}\footnote{\ \textcolor{purple}{Specifically, we do not assume regions of $\covariateSpace$ with constant $\individualEffect$ (e.g., a step-function data-generating process for treatment effects).}} 

\textcolor{purple}{The task of GATE analysis is therefore to form groups in a principled way---deciding which covariates define the grouping and how many groups $G$ to report---and to obtain valid inference for the resulting GATEs.\footnote{\ \textcolor{purple}{While higher values of $G$ can uncover more detailed heterogeneity, partitions that are too fine may not offer substantial advantages in terms of interpretability compared to CATEs.}} This paper proposes a data-driven procedure that constructs partitions of $\covariateSpace$ at various levels of granularity and shows how to obtain valid inference for the group effects.}

%%% IDENTIFICATION.
\subsection{Identification}
\label{subsec_framework_identification}
\noindent All the estimands discussed in the previous section are defined in terms of potential outcomes. However, each unit is either treated or not treated. We thus observe only one potential outcome per unit, and further assumptions are needed for identification.

The following standard assumptions are sufficient to identify the CATEs \parencite[see, e.g.,][]{imbens2015causal}:

\begin{assumption}
    (Unconfoundedness): $\{ \potentialOutcome{0}, \potentialOutcome{1} \} \indep D_i | X_i$.
    \label{assumption_unconfoundedness}
\end{assumption}

\begin{assumption}
    (Common support): $0 < \propensityScore{X_i} < 1$, where $\propensityScore{X_i} := \PX ( D_i = 1 | X_i )$ is the conditional treatment probability (or propensity score).
    \label{assumption_common_support}
\end{assumption}

\noindent The unconfoundedness assumption requires that $X_i$ contains all \open confounder" jointly affecting the treatment assignment and the outcome.\footnote{\ $X_i$ can also include additional \open heterogeneity covariates" not necessary for identification but for which effect heterogeneity is of interest. The sets of confounders and heterogeneity covariates can overlap in any way or be disjoint.} The common support assumption states that each unit must have a non-zero probability of belonging to the treatment and control groups.

Under Assumptions \ref{assumption_unconfoundedness}--\ref{assumption_common_support}, the CATEs are identified from observable data:

\vspace{2pt}

\begin{equation}
    \begin{split}
        \conditionalMean{1}{X_i} & - \conditionalMean{0}{X_i} \\
        & = \EX [ \potentialOutcome{1} | D_i = 1, X_i  ] - \EX [ \potentialOutcome{0} | D_i = 0, X_i  ] \,\,\,\,\,\,\, (\textit{by equation \ref{equation_sutva}}) \\
        & = \EX [ \potentialOutcome{1} | X_i  ] - \EX [ \potentialOutcome{0} | X_i  ] \,\,\,\,\,\,\,\,\,\,\,\,\,\,\,\,\,\,\,\,\,\,\,\,\,\,\,\,\,\,\,\,\,\,\,\,\,\,\,\,\,\,\,\,\, (\textit{by Assumption \ref{assumption_unconfoundedness}}) \\
        & = \tau ( X_i ),
    \end{split}
    \label{equation_identification}
\end{equation}
where $\conditionalMean{D_i}{X_i} := \EX [ Y_i | D_i, X_i  ]$ and Assumption \ref{assumption_common_support} ensures that the conditional expectations are well-defined for all values within the support of $X_i$. The ATE and GATEs are expressed as expectations of the CATEs and are thus identified under the same assumptions.

%%%%%%%%%%%%%%%%%%%%%%%%%%%%%%%%%%%%%%%
%          AGGREGATION TREES.         %
%%%%%%%%%%%%%%%%%%%%%%%%%%%%%%%%%%%%%%%

\section{Aggregation trees}
\label{sec_aggregation_trees}
\noindent This section outlines the implementation of the methodology proposed in this paper, which involves three steps. First, an estimation step constructs an estimate $\hat{\tau} ( \cdot )$ of $\tau ( \cdot )$. Second, a tree-growing step approximates $\hat{\tau} ( X_i )$ by a standard decision tree \parencite{breiman1984cart} that constrains the set of admissible groupings. Third, a tree-pruning step generates a sequence of nested subtrees, one for each level of granularity. Each subtree provides an optimal grouping, where optimality means that, at each granularity level, the groupings minimize the loss in explained heterogeneity due to aggregation.

The next subsection describes the tree-growing step, detailing the splitting strategy used by aggregation trees and comparing it with that of causal trees \parencite{athey2016recursive}. Then, the tree-pruning step is discussed. Finally, we explain how to conduct valid inference about the GATEs using double machine learning procedures. \textcolor{purple}{The algorithm below summarizes the full implementation.}

\begin{algorithm}[b]
  \footnotesize
  \captionsetup{labelformat=empty}
  \caption{\textbf{Algorithm}\quad Aggregation trees}
  {\setstretch{1.20}
  \textbf{Inputs:} Outcome vector $Y$, (binary) treatment vector $D$, and covariate matrix $X$.\\
  \newlength{\outwd}\settowidth{\outwd}{\textbf{Outputs:} }
  \textbf{Outputs:} (i) Sequence of partitions $\tree_{\alpha_{0}}, \tree_{\alpha_{1}}, \dots, \tree_{\alpha_{\max}}$\\
  \hspace*{\outwd}(ii) GATE estimates $\{\hat{\tau}_g\}_{g=1}^{|\tree_{\alpha}|}$ and their standard errors for each partition $\tree_{\alpha}$.

  \textbf{Procedure:}

  {\algrenewcommand\alglinenumber[1]{}
  \begin{algorithmic}
    \Statex \quad $(Y_{\mathrm{tr}}, D_{\mathrm{tr}}, X_{\mathrm{tr}}), (Y_{\mathrm{hon}}, D_{\mathrm{hon}}, X_{\mathrm{hon}}) \leftarrow \texttt{SampleSplit}(Y, D, X)$ \Comment{training/honest split}
    
    \Statex \textbf{i. Constructing sequence of groupings}
    \Statex \quad $\hat{\tau} (\cdot) \leftarrow \texttt{EstimateCATE}(Y_{\mathrm{tr}}, D_{\mathrm{tr}}, X_{\mathrm{tr}})$  \Comment{e.g., causal forest, X-learner}
    \Statex \quad $\mathcal{T}_{0} \leftarrow \texttt{GrowTreeCART}(\hat{\tau} (\cdot), X_{\mathrm{tr}})$ \Comment{e.g., \texttt{rpart}}
    \Statex \quad $\{ \tree_{\alpha_k} \}_{k = 0}^{\max} \leftarrow \texttt{PruningTree}(\mathcal{T}_{0})$ \Comment{cost-complexity pruning}

    \Statex \textbf{ii. Estimation and inference}
    \Statex \quad $\tree_{\alpha^*} \leftarrow \texttt{SelectPartition}(\{ \tree_{\alpha_k} \}_{k = 0}^{\max})$ \Comment{e.g., policy relevance or cross-validation}
    \Statex \quad $\{ \hat{\tau}_g \}_{g = 1}^{|\tree_{\alpha^*}|} \leftarrow \texttt{EstimateGATEs}(\tree_{\alpha^*}, Y_{\mathrm{hon}}, D_{\mathrm{hon}}, X_{\mathrm{hon}} )$ \Comment{OLS/DR regression on leaf dummies}
  \end{algorithmic}
  }
  }
\end{algorithm}

%%% TREE GROWING.
\subsection{Tree-growing step}
\label{subsec_treegrowing}
\noindent Trees are typically constructed by greedily minimizing an assumed loss function based on the mean squared error criterion. This minimization follows the approach proposed by \textcite{breiman1984cart}, who suggest recursively stratifying the covariate space using axis-aligned splits.

Starting with a region of the covariate space $\mathcal{R}_m \subseteq \covariateSpace$, consider a candidate splitting variable $X_{ij}$ and splitting point $x$. We define the corresponding subregions as:\footnote{\ In the case of categorical splitting variables, $x$ corresponds to a subset of possible levels of $X_{ij}$, and the inequality signs are replaced by $\in$ and $\notin$.}
\begin{equation}
    \begin{gathered}
        \mathcal{R}_{m + 1} ( j, x ) = \{ X_i | X_{ij} \leq x \}, \quad \mathcal{R}_{m + 2} ( j, x ) = \{ X_i | X_{ij} > x \}.
    \end{gathered}
\end{equation}
\noindent The split occurs on some pair $( j, x )$, and the population is stratified accordingly. The process is then repeated in the resulting subregions, thus obtaining increasingly finer partitions of $\covariateSpace$. The whole procedure can be described by the shape of a decision tree: the \open root” (i.e., the node with no \open parent") corresponds to $\covariateSpace$, the $m$-th internal node represents subregion $\mathcal{R}_m$ and has two \open children" nodes representing subregions $\mathcal{R}_{m + 1}$ and $\mathcal{R}_{m + 2}$, and the \open leaves" (i.e., the collection of terminal nodes) correspond to a partition of $\covariateSpace$.

Ideally, we would like to explore the space of all possible trees and pick the one whose associated partition minimizes the assumed loss function. However, it is generally infeasible to enumerate all the distinct binary decision trees.\footnote{\ Consider the situation where $X_i$ is composed of $p$ binary covariates, and let $\mathcal{D}$ be the \open depth" of a given tree (i.e., the number of nodes connecting the root to the furthest leaf). Appendix \ref{app_n_trees} shows that $L_{\mathcal{D}} = \prod_{d = 1}^\mathcal{D} ( p - ( d - 1 ) )^{2^{d - 1}}$ is a lower bound for the number of distinct binary decision trees grown by recursively partitioning $\covariateSpace$ and having a depth equal to or lower than $\mathcal{D}$. $L_{\mathcal{D}}$ quickly diverges as $p$ grows. For example, fixing $\mathcal{D} = 3$ and letting $p = 10$ yields a lower bound of 3,317,760, while letting $p = 20$ leads to a bound of 757,926,720. Things only worsen with categorical covariates taking more than two values or continuous covariates discretized using a large number of bins.} To cope with this issue, \textcite{breiman1984cart} suggest a \open greedy" approach that partitions each region $\mathcal{R}_m \subseteq \covariateSpace$ by choosing the split that minimizes the assumed loss function within the resulting subregions $\mathcal{R}_{m + 1}$ and $\mathcal{R}_{m + 2}$. This process is then iterated until some particular \open stopping criterion" is met, for instance the maximum depth of the tree. This approach is greedy in that it ignores that a suboptimal split could yield better results at later steps and is generally considered to be a reasonable way of circumventing the exhaustive search of the space of all possible trees. 

Let $\tree$ be some tree constructed using a training sample $\trainingSample$, and let $\testSample$ be an independent test sample. Then, when heterogeneous treatment effects are the object of the analysis, one wants to build a tree that minimizes $EMSE ( \tree ) = \EX [ \mseCriterion{ \testSample}{\trainingSample}{\tree} ]$, where the expectation is taken over the joint distribution of the training and test samples and:\footnote{\ We are departing from the standard criterion $\EX [ \{ \tau_i - \tilde{\tau} ( X_i, \trainingSample, \tree ) \}^2 ]$ by subtracting $\EX [ \tau_i^2 ]$. Because this term does not depend on an estimator, the tree that minimizes the standard criterion also minimizes $EMSE ( \cdot )$.}
\begin{equation}
   \begin{split}
        MSE ( \testSample, \trainingSample, \tree ) & = \frac{1}{| \testSample |} \sum_{i \in \testSample } \{ [ \tau_i - \tilde{\tau} ( X_i, \trainingSample, \tree ) ]^2 - \tau_i^2 \} \\
        & = \frac{1}{| \testSample |} \sum_{i \in \testSample } \tilde{\tau}^2 ( X_i, \trainingSample, \tree ) - \frac{2}{| \testSample |} \sum_{i \in \testSample } \tau_i \, \tilde{\tau} ( X_i, \trainingSample, \tree ),
    \end{split} 
\end{equation}
\noindent with $\tau_i \equiv \tau ( X_i )$ and $\tilde{\tau} ( x, \mathcal{S}, \tree )$ some estimate of $\cate{\cdot}$ within the leaf $\ell ( x, \tree )$ of $\tree$ where $x$ falls obtained using observations in the sample $\mathcal{S}$. In practice, trees are constructed by greedily minimizing an in-sample loss function $MSE ( \trainingSample, \trainingSample, \tree )$.\footnote{\ This is what \textcite{athey2016recursive} denote as the \open adaptive" case, where the same sample is used to both construct and estimate the tree. \textcite{athey2016recursive} also consider an alternative \open honest" criterion $MSE ( \testSample, \honestSample, \tree )$ that uses different samples for construction of the tree ($\trainingSample$) and treatment effect estimation ($\honestSample$). For simplicity, we focus on the adaptive case here and postpone the discussion of honesty to a later section.}

The key challenge in a causal inference framework is that we do not observe $\tau_i$. Thus, $MSE ( \cdot, \cdot, \cdot )$ is an infeasible criterion and needs to be estimated. We propose an estimator of $MSE ( \cdot, \cdot, \cdot )$ obtained by plugging an estimate $\estimatedCate{\cdot}$ of $\cate{\cdot}$ constructed from the training sample in a preliminary estimation step:
\begin{equation}
    \widehat{MSE}_{\scriptstyle_{AT}} ( \testSample, \trainingSample, \tree ) = \frac{1}{| \testSample |} \sum_{i \in \testSample} \tilde{\tau}_{\scriptstyle_{AT}}^2 ( X_i, \trainingSample, \tree ) - \frac{2}{| \testSample |} \sum_{i \in \testSample} \hat{\tau}_i \, \tilde{\tau}_{\scriptstyle_{AT}} ( X_i, \trainingSample, \tree ),
\end{equation}
\noindent with:
\begin{equation}
    \tilde{\tau}_{\scriptstyle_{AT}} ( x, \mathcal{S}, \tree ) = \frac{1}{| i \in \mathcal{S} : X_i \in \ell ( x, \tree ) |} \sum_{i \in \mathcal{S} : X_i \in \ell ( x, \tree ) } \hat{\tau}_i.
\end{equation}
\noindent If we use a consistent estimator of $\tau_i$, then $\widehat{MSE}_{\scriptstyle_{AT}} ( \testSample, \trainingSample, \tree )$ is an approximately unbiased estimator of $MSE ( \testSample, \trainingSample, \tree )$ if the assignment to treatment is random conditional on $X_i$. In practice, we construct aggregation trees by greedily minimizing the in-sample loss function $\widehat{MSE}_{\scriptstyle_{AT}} ( \trainingSample, \trainingSample, \tree )$. This is equivalent to selecting splits that minimize the conditional variance of $\hat{\tau}_i$ in the resulting nodes. Consequently, the greedy approach partitions each region $\mathcal{R}_m \subseteq \covariateSpace$ in a way that maximizes systematic heterogeneity between the resulting subgroups, thereby constructing a set of admissible groupings represented by the resulting tree $\tree_0$.

\textcite{athey2016recursive} propose instead the following estimator of $MSE ( \cdot, \cdot, \cdot )$:
\begin{equation}
    \widehat{MSE}_{\scriptstyle_{CT}} ( \testSample, \trainingSample, \tree ) = \frac{1}{| \testSample |} \sum_{i \in \testSample} \tilde{\tau}_{\scriptstyle_{CT}}^2 ( X_i, \trainingSample, \tree ) - \frac{2}{| \testSample |} \sum_{i \in \testSample} \tilde{\tau}_{\scriptstyle_{CT}} ( X_i, \testSample, \tree ) \tilde{\tau}_{\scriptstyle_{CT}} ( X_i, \trainingSample, \tree ),
\end{equation}
\noindent with:
\begin{equation}
    \tilde{\tau}_{\scriptstyle_{CT}} ( x, \mathcal{S}, \tree ) = \hat{\mu} (1, x, \mathcal{S}, \tree ) - \hat{\mu} (0, x, \mathcal{S}, \tree ),
\end{equation}
\noindent and
\begin{equation}
    \hat{\mu} ( d, x, \mathcal{S}, \tree ) = \frac{1}{| i \in \mathcal{S} : X_i \in \ell ( x, \tree ), D_i = d |} \sum_{i \in \mathcal{S} : X_i \in \ell ( x, \tree ), D_i = d} Y_i.
\end{equation}
\noindent In randomized experiments, $\widehat{MSE}_{\scriptstyle_{CT}} ( \testSample, \trainingSample, \tree )$ is an approximately unbiased estimator of $MSE ( \testSample, \trainingSample, \tree )$, as $\EX [ \tau_i | i \in \testSample : i \in\ell ( x, \tree ) ] = \EX [ \tilde{\tau}_{\scriptstyle_{CT}} ( x, \testSample, \tree ) ]$, with the expectations taken over the distribution of the test samples. As before, we construct causal trees by greedily minimizing the in-sample loss function $\widehat{MSE}_{\scriptstyle_{CT}} ( \trainingSample, \trainingSample, \tree )$.

We expect the splitting strategy of aggregation trees to result in a lower sampling variance, especially when dealing with covariates that influence outcome levels but not treatment effects. For example, consider potential outcomes expressed as
\begin{equation}
    Y_i ( d ) = \phi ( X_i ) + \frac{1}{2} ( 2 d - 1 ) \tau ( X_i ) + \epsilon_i,
\end{equation}
with $\phi ( X_i ) = \frac{1}{2} X_{i1} + X_{i2}$ a model for the mean effect and $\tau ( X_i ) = \frac{1}{2} X_{i1}$. When exploring covariate values as potential splitting points, we shift one observation at a time from one region of the covariate space to its complement. Since each observation belongs to either the treatment or control group, this alters the sample average of the observed outcomes of only one group, thus affecting $\tilde{\tau}_{\scriptstyle_{CT}} ( \cdot, \cdot, \cdot )$. Due to the substantial influence of $X_{i2}$ on mean outcomes, moving a single observation between child nodes based on this covariate can substantially alter the sample average of the observed outcomes of one group. Consequently, we expect $\tilde{\tau}_{\scriptstyle_{CT}} ( \cdot, \cdot, \cdot )$ to exhibit considerable variability with the choice of splitting point, even though $X_{i2}$ does not enter the model for $\cate{\cdot}$. This variability may also cause the estimator to identify spurious splits involving $X_{i2}$. In contrast, for accurately estimated CATEs, $\tilde{\tau}_{\scriptstyle_{AT}} ( \cdot, \cdot, \cdot )$ is expected to remain stable when shifting a single observation between child nodes based on $X_{i2}$. Consequently, we expect that $\tilde{\tau}_{\scriptstyle_{AT}} ( \cdot, \cdot, \cdot )$ will vary less with the choice of splitting point, resulting in a lower sampling variance.

%%% TREE PRUNING STEP.
\subsection{Tree-pruning step}
\label{subsec_treepruning}
\noindent After a deep tree has been constructed, the standard practice is to prune it according to an assumed cost-complexity criterion. Aggregation and causal trees rely on the same criterion, which is composed of two terms:
\begin{equation}
    C_{\alpha} ( \tree ) = \mseCriterion{\trainingSample}{\trainingSample}{\tree} + \alpha |\tree|.
    \label{equation_pruning_criterion}
\end{equation}
\noindent The first term corresponds to the loss function used for constructing the tree and measures the in-sample goodness-of-fit of the model. The second term is a regularization component that penalizes the model's complexity---defined as the number of leaves $|\tree|$---according to the cost-complexity parameter $\alpha \in [ 0, \infty )$. Regularization is needed to prevent overfitting: the in-sample loss function $MSE ( \mathcal{S}^{tr}, \mathcal{S}^{tr}, \tree )$ always decreases with additional splits, even in the cases where the out-of-sample $MSE ( \mathcal{S}^{te}, \mathcal{S}^{tr}, \tree )$ actually increases.

The parameter $\alpha$ controls the relative weight of the two components and thus the balance between the accuracy and the interpretability of the model. Define a subtree $\tree \subset \tree_0$ as any tree that can be obtained by collapsing any number of internal nodes of $\tree_0$ and let $\tree_{\alpha} \subseteq \tree_0$ be the smallest subtree for which (\ref{equation_pruning_criterion}) is minimized. For each $\alpha$, a unique $\tree_{\alpha}$ exists, which can be identified by \open weakest link pruning”: starting from $\tree_0$, we iteratively collapse the internal node that gives the slightest increase in the accuracy of the approximation.

Following this procedure, we can generate a sequence of nested subtrees $\tree_{\alpha_{\scriptstyle_{0}}}, \tree_{\alpha_{\scriptstyle_{1}}} \dots, \tree_{\alpha_{\scriptstyle_{max}}}$, where $0 = \alpha_{\scriptstyle_{0}} < \alpha_{\scriptstyle_{1}} < \dots < \alpha_{\scriptstyle_{max}} < \infty$ are threshold values such that all $\alpha$ in a given interval lead to the same subtree and $\tree_{\alpha_{\scriptstyle_{max}}}$ corresponds to the tree's root. Each subtree in this sequence is associated with a partition of the covariate space. Therefore, the tree-pruning step generates a sequence of groupings, one for each threshold value $\alpha_{\scriptstyle_{0}} < \alpha_{\scriptstyle_{1}} < \dots < \alpha_{\scriptstyle_{max}}$. 

Cross-validation procedures are commonly used to determine the optimal cost-complexity parameter and thus select a single partition \parencite[see, e.g.,][]{hastie2009elements}.\footnote{\ \textcite{athey2016recursive} also utilize cross-validation, adapting the standard criterion for the honest case.} However, the whole sequence of groupings generated by the pruning process is also of significant interest. Because each tree $\tree_{\alpha_{\scriptstyle_{k}}}$ is obtained by collapsing the weakest node of $\tree_{\alpha_{\scriptstyle_{k - 1}}}$, the tree-pruning step constructs optimal groupings by aggregating the two subgroups for which the loss in explained heterogeneity resulting from aggregation is minimized. Moreover, because the sequence is nested, subgroups formed at a particular level of granularity are never disrupted at coarser levels. This property guarantees the consistency of the results across the different granularity levels \parencite[see, e.g.,][]{cotterman1992classification}. Consequently, the sequence allows for a flexible and coherent exploration of the trade-off between interpretability and the discovery of more granular heterogeneity.\footnote{\ A single partition can still be selected for practical purposes using standard cross-validation methods.}

%% ESTIMATION AND INFERENCE.
\subsection{Estimation and inference}
\label{subsec_estimation_inference}
\noindent For a particular grouping $\tree_{\alpha}$, we can estimate the GATEs in several ways. In randomized experiments, taking the difference between the mean outcomes of treated and control units in each group is an unbiased estimator of the GATEs. Equivalently, we can obtain the same point estimates in addition to their standard errors by estimating via OLS the following linear model:
\begin{equation}
    Y_i = \sum_{l = 1}^{|\mathcal{T_{\alpha}}|} L_{i, l} \, \gamma_l + \sum_{l = 1}^{|\tree_{\alpha}|} L_{i, l} \, D_i \, \beta_l + \epsilon_i,
    \label{equation_randomized_experiments_ols}
\end{equation}
with $L_{i, l}$ a dummy variable equal to one if the $i$-th unit falls in the $l$-th leaf of $\tree_{\alpha}$. Exploiting the random assignment to treatment, we can show that each $\beta_l$ identifies the GATE in the $l$-th leaf. 

In observational studies, estimating model (\ref{equation_randomized_experiments_ols}) would yield biased GATE estimates due to the selection into treatment. To get unbiased estimates, we can use the orthogonal estimator of \textcite{semenova2021debiased} to estimate the best linear predictor of $\cate{\cdot}$ given a set of dummies denoting leaf membership.\footnote{\ \textcolor{purple}{One can also conduct sensitivity analysis to assess robustness to violations of Assumption \ref{assumption_unconfoundedness} \parencite[see, e.g.,][]{lee2021discovering}.}} The key idea is to construct a random variable $\score_i$, generally called score, such that $\tau ( X_i ) = \EX [ \score_i | X_i ]$, and project it onto $L_{i, 1}, \dots, L_{i, |\tree_{\alpha}|}$. 

Consider the doubly-robust scores of \textcite{robins1995semiparametric}:
\begin{equation}
    \score_i^{DR} = \conditionalMean{1}{X_i} - \conditionalMean{0}{X_i} + \frac{D_i [ Y_i - \conditionalMean{1}{X_i} ]}{\propensityScore{X_i}}  - \frac{ ( 1 - D_i ) [ Y_i - \conditionalMean{0}{X_i} ]}{1 - \propensityScore{X_i}}.
    \label{equation_dr_scores}
\end{equation} 
Because $\EX [ \score_i^{DR} | X_i ] = \tau ( X_i )$, this score is a natural candidate. We recognize that it depends on unknown functions $\eta ( X_i ) := \{ \conditionalMean{1}{X_i}, \conditionalMean{0}{X_i}, \propensityScore{X_i} \}$ and make this explicit by writing $\score_i^{DR} := \score_i^{DR} ( \eta )$. We refer to $\eta$ as nuisance functions, as they are not of direct interest but necessary to construct a plug-in estimate $\score_i^{DR} ( \hat{\eta} )$ of $\score_i^{DR} ( \eta )$ that we aim to regress on $L_{i, 1}, \dots, L_{i, |\tree_{\alpha}|}$. 

\textcite{semenova2021debiased} show that $\score_i^{DR} ( \eta )$ is a Neyman-orthogonal score \parencite{chernozhukov2018double}, that is, its plug-in estimate $\score_i^{DR} ( \hat{\eta} )$ is insensitive to bias in the estimation of $\hat{\eta}$. They then suggest the following two-stage procedure. First, construct an estimate $\hat{\eta}$ of the nuisance functions $\eta$ using $K$-fold cross-fitting: split the sample into $K$ folds of similar sizes and, for each $k = 1, \dots, K$, estimate $\hat{\eta}_k$ using all but the $k$-th folds. Second, construct $\widehat{\score}_i^{DR} := \score_i^{DR} ( \hat{\eta}_k )$, where the observation $i$ belongs to the $k$-th fold, and estimate via OLS the following linear model:
\begin{equation}
    \widehat{\score}_i^{DR} = \sum_{l = 1}^{|\tree_{\alpha}|} L_{i, l} \, \beta_l + \epsilon_i.
    \label{equation_observational_studies_ols}
\end{equation}
As before, each $\beta_l$ identifies the GATE in the $l$-th leaf. Moreover, \textcite{semenova2021debiased} show that thanks to the Neyman-orthogonality of $\score_i^{DR}$, the OLS estimator $\hat{\beta}_l$ of $\beta_l$ is root-$n$ consistent and asymptotically normal, provided that the product of the convergence rates of the estimators of the nuisance functions $\conditionalMean{\cdot}{\cdot}$ and $\propensityScore{\cdot}$ is faster than $n^{1/2}$. This allows using machine learning estimators such as random forests and LASSO to estimate the nuisance functions, as they are shown to achieve an $n^{1/4}$ convergence rate and faster under particular conditions.

However, GATE estimates may show some bias if we use the same data to construct the tree and to estimate (\ref{equation_randomized_experiments_ols})--(\ref{equation_observational_studies_ols}), leading to invalid inference. One way out is to grow \open honest" trees \parencite{athey2016recursive}. Honesty is a subsample-splitting technique that requires that different observations are used to form the subgroups and estimate the GATEs. For this purpose, we split the observed sample into a training sample $\trainingSample$ and an honest sample $\honestSample$ of arbitrary sizes. We use $\trainingSample$ to estimate $\cate{\cdot}$ and construct the sequence of groupings and, for a particular grouping $\tree_{\alpha}$, we use $\honestSample$ to estimate (\ref{equation_randomized_experiments_ols})--(\ref{equation_observational_studies_ols}). This way, the asymptotic properties of GATE estimates are the same as if the groupings had been exogenously given. Therefore, we can use the estimated standard errors to conduct valid inference as usual, e.g., by constructing conventional confidence intervals.\footnote{\ \textcolor{purple}{When the number of reported groups $|\tree_{\alpha}|$ is large, it is prudent to account for multiplicity across GATEs \parencite[see, e.g.,][]{bargagli2022heterogeneous}, for example by adjusting $p$-values to account for multiple hypothesis testing controlling the familywise error rate \parencite[see, e.g.,][]{holm1979multiple, hochberg1988sharper, hommel1988stagewise, romano2005exact} or the false discovery rate \parencite[see, e.g.,][]{benjamini1995controlling, benjamini2001control}. If the partition becomes so fine that interpretability is compromised, an alternative is to summarize heterogeneity along a low-dimensional effect modifier---essentially shifting the estimand from GATEs to a \open reduced-dimensional" CATE. \textcite{fan2022estimation} provide estimators and uniform inference procedures for the reduced-dimensional CATE.}} However, honesty generally comes at the expense of a larger mean squared error, as fewer observations are used to estimate $\estimatedCate{\cdot}$, construct the tree, and compute GATE estimates.

%%%%%%%%%%%%%%%%%%%%%%%%%%%%%%%%%%%%%%%%%%%%%%
%            EMPIRICAL MONTE-CARLO.          %
%%%%%%%%%%%%%%%%%%%%%%%%%%%%%%%%%%%%%%%%%%%%%%

\section{Empirical Monte-Carlo}
\label{sec_empirical_montecarlo}
\noindent We conduct an empirical Monte-Carlo study \parencite{huber2013performance, lechner2013sensitivity} to investigate the performance of aggregation and causal trees in a context that closely approximates a real application.\footnote{\ See also \textcite{lechner2018modified, knaus2021machine, lechner2022modified} for more recent implementations of empirical Monte-Carlo studies.} Empirical Monte-Carlo studies aim to base the data-generating processes (DGPs) on real data as much as possible, thereby evaluating the estimators in a realistically representative context. The idea is to use a sufficiently large data set as the population of interest, from which we can draw random samples for estimation.\footnote{\ \textcolor{purple}{We do not benchmark aggregation trees against causal forests \parencite{wager2018estimation} or other causal machine learning methods \parencite[see, e.g.,][]{athey2019generalized, kunzel2019metalearners, lechner2022modified} because the targets differ: those approaches estimate unit-level CATEs, whereas aggregation trees construct interpretable groupings and estimate the corresponding GATEs. Causal trees return an explicit partition with leaf-level GATEs and are therefore the natural benchmark.}}

In the following subsections, we detail the implementation of the empirical Monte-Carlo study. First, we describe the data set forming our population. Next, we illustrate the DGPs, the implementation of estimators, and the performance measures we employ. Finally, we present the results.

\subsection{Population}
\label{subsec_empirical_montecarlo_population}
\noindent We base our simulations on a data set previously used to evaluate the impact of maternal smoking on birth weight \parencite[see, e.g.,][]{almond2005costs, cattaneo2010efficient, heiler2021effect, bodory2022high}. Since we will use the same data set in our empirical illustration in the next section, we postpone the discussion of previous findings to that section and focus here on the details of the data.

The clean data set consists of $435{,}124$ observations measured in Pennsylvania between $1989$ and $1991$. The outcome of interest is the infant's weight at birth in grams. The treatment indicator equals one if the mother smoked during pregnancy and zero otherwise. The pre-treatment covariate vector contains $39$ confounders and heterogeneity variables, providing information on the mother's and father's background characteristics (age, ethnicity, whether the mother was married or foreign-born), mother's behavior possibly associated with smoking (whether she drank alcohol during pregnancy, how many drinks per week), maternal medical risk factors not affected by smoking during pregnancy, and birth characteristics (e.g., whether the infant is first born, number and quality of prenatal care visits).\footnote{\ Table \ref{table_data_description} in Appendix \ref{app_data_description} provides a description of all the variables.}

To avoid common support issues, we drop children whose parents were particularly young or old at birth, or who attended more than thirty prenatal care visits. Moreover, we drop children whose mothers used to consume more than ten alcoholic drinks per week during pregnancy. A total of $596$ observations are removed from the original data set.

Table \ref{table_descriptive_stats} presents the summary statistics for the treated and control groups in the final sample. The table displays sample averages and standard deviations for each variable, along with two measures of difference in the distribution across treatment arms: the normalized difference, measuring the difference between the locations of the distributions, and the logarithm of the ratio of standard deviations, measuring the difference in the dispersion of the distributions. Overall, the sample appears to be sufficiently balanced, with only five relatively unbalanced covariates: \texttt{meduc}, \texttt{unmarried} and \texttt{feduc} exhibit strong differences in locations, while \texttt{alcohol} and \texttt{n\_drink} exhibit strong differences in dispersion. These results are robust to the inclusion of the excluded observations.

%%% SIMULATION DETAILS.
\subsection{Simulation details}
\label{subsec_empirical_montecarlo_details}
\noindent We estimate the propensity score via a logistic regression using the full sample and all baseline covariates. The resulting estimate $\estimatedPropensityScore{\cdot}$ is later used as the true selection model in the simulation to ensure that the selection behavior closely mimics that observed in the real data. We then remove all treated units from the sample. This implies that we observe $\potentialOutcome{0}$ for all units. 

We now need to specify a model for the individual effects. Since model specification can be somewhat arbitrary, we aim to reduce this arbitrariness and better approximate a realistic scenario by fitting an honest causal forest \parencite{athey2019generalized} using the full sample and choosing a model that mimics the resulting predictions.\footnote{\ While it would be possible to directly use the estimated CATEs in the DGPs, this approach might bias our results towards estimators similar to those used for CATE estimation \parencite{knaus2021machine}.} We find that all predicted CATEs are negative and statistically different from zero at the $5\%$ significance level. Therefore, we specify the following model for the individual effects:\footnote{\ Figure \ref{fig_ites_cates} in Appendix \ref{app_further_simulation_results} displays the estimated CATEs sorted by magnitude alongside the individual effects generated by model (\ref{equation_model_individual_effects}).} 
\begin{equation}
    \xi ( X_i ) = - a \normalPdf{\normalizedPropensityScore{X_i}} - \max_i \{ - a \normalPdf{\normalizedPropensityScore{X_i}} \},
    \label{equation_model_individual_effects}
\end{equation}
with $\normalizedPropensityScore{X_i} = \frac{\estimatedPropensityScore{X_i}}{\max_{i} \estimatedPropensityScore{X_i}}$ a normalized version of the estimated propensity score $\estimatedPropensityScore{X_i}$, and $\normalPdf{\cdot}$ the standard normal probability distribution function. The parameter $a$ controls the degree of heterogeneity. If $a = 0$, all individual effects are zero. As $a$ increases, the degree of heterogeneity increases as well.

We use model (\ref{equation_model_individual_effects}) to compute $\estimatedPotentialOutcome{1} = \potentialOutcome{0} + \xi ( X_i )$ for all units. We then set aside a random validation sample of $10{,}000$ units to evaluate our performance measures detailed below. This approach allows us to assess the out-of-sample predictive power of estimators under investigation \parencite{knaus2021machine}. We treat the remaining $343{,}140$ units as our population from which we draw random samples for estimation.

After drawing a sample of size $n$, we assign the treatment using a Bernoulli process. We consider two scenarios: one with random assignment ($D_i \sim \textit{Bernoulli} ( 0.5 )$) and another with assignment based on the \open true" propensity score ($D_i \sim \textit{Bernoulli} ( \estimatedPropensityScore{X_i} )$). Notice that in the latter case $\estimatedPropensityScore{\cdot}$ is used in both the model for individual effects and for treatment assignment. This complicates the task for estimators to separate selection bias from effect heterogeneity. Finally, we construct the observed outcomes for units in the drawn sample as in (\ref{equation_sutva}).

We split each drawn sample into a training sample $\trainingSample$ and an honest sample $\honestSample$ of equal sizes. We then construct a causal tree and two aggregation trees.\footnote{\ \textcolor{purple}{Aggregation trees are constructed with the R package \texttt{rpart} using default settings. Causal trees are constructed with the R package \texttt{causalTree} using default settings, except for the \texttt{minsize} parameter---which controls the minimum number of treated and control units required in a leaf to attempt a split---which we set to $4$ (default is $2$) to reduce the risk of \open empty-arm" leaves in the honest sample (i.e., leaves that, when populated with observations from the honest sample, contain zero treated or control units).}} To build the aggregation trees, we estimate $\cate{\cdot}$ using the X-learner \parencite{kunzel2019metalearners} and the causal forest \parencite{athey2019generalized} estimators. We use standard cross-validation procedures to select a single partition from the resulting causal and aggregation trees.\footnote{\ \textcolor{purple}{For causal trees, we use the honest cross‐validation criterion of \textcite{athey2016recursive}. For aggregation trees, we use the standard criterion.}} All these operations are performed utilizing only observations from $\trainingSample$.

To obtain point estimates and standard errors for the GATEs, causal trees follow the approach of \textcite{athey2016recursive} and estimate model (\ref{equation_randomized_experiments_ols}). On the other hand, aggregation trees estimate model (\ref{equation_observational_studies_ols}). Honest regression forests and $5$-fold cross-fitting are employed to estimate the nuisance functions necessary for constructing the doubly-robust scores $\score_i^{DR}$. All these operations are performed using only observations from $\honestSample$.

We use the external validation sample to assess the quality of estimation. Three performance measures are computed: the root mean squared error, the absolute bias, and the standard deviation of the predictions for each observation in the validation sample:\footnote{\ Note that we are evaluating which estimator best approximates the individual effects $\xi ( \cdot )$. However, the relative performance of the estimators in approximating the unknown CATEs is the same, as the estimator that minimizes the mean squared error for $\xi ( x )$ also minimizes the mean squared error for $\cate{x}$ \parencite{kunzel2019metalearners}.}
\begin{equation}
    \begin{gathered}
        RMSE ( x ) = \sqrt{\frac{1}{R} \sum_{r = 1}^R [ \hat{\tau}_r ( x ) - \xi ( x ) ]^2}, \\
        \vert Bias ( x ) \vert = \left| \frac{1}{R} \sum_{r = 1}^R \hat{\tau}_r ( x ) - \xi ( x ) \right|, \quad SD ( x ) = \sqrt{\frac{1}{R} \sum_{r = 1}^R [ \hat{\tau}_r ( x ) - \frac{1}{R} \sum_{r = 1}^R \hat{\tau}_r ( x ) ]^2},
    \end{gathered}
\end{equation}
with $x$ a generic point in the validation sample, $R$ the number of replications, and $\hat{\tau}_r ( \cdot )$ the CATE estimated by the tree at the $r$-th replication. We summarize these performance measures by averaging over the validation sample. Additionally, we evaluate the actual coverage rates of conventional $95\%$ confidence intervals for the GATEs constructed using the estimated standard errors.

%%% RESULTS.
\subsection{Results}
\label{subsec_empirical_montecarlo_results}
\noindent Table \ref{table_simulation_results_honest} presents the results obtained from $R = 1{,}000$ replications across three sample sizes and two levels of heterogeneity: low ($a = 20$) and high ($a = 50$). Overall, aggregation trees outperform causal trees in terms of prediction accuracy, with both $AT_{\scriptstyle_{XL}}$ and $AT_{\scriptstyle_{CF}}$ showing lower RMSE than causal trees. This effect is particularly strong when treatment is randomly assigned, where the RMSE of causal trees is between $28\%$ and $121\%$ larger than that of aggregation trees. When treatment assignment is based on $\estimatedPropensityScore{\cdot}$, the RMSE of causal trees is still larger by $17\%$ to $66\%$, except for the smallest sample size, where causal trees marginally outperform $AT_{\scriptstyle_{CF}}$.  $AT_{\scriptstyle_{XL}}$ and $AT_{\scriptstyle_{CF}}$ consistently show similar perfomances. 

\begingroup
  \setlength{\tabcolsep}{8pt}
  \renewcommand{\arraystretch}{1.1}
  \begin{table}[b!]
     \centering
     \begin{adjustbox}{width = 1\textwidth}
     \begin{tabular}{@{\extracolsep{5pt}}l c c c c c c c c c c c c }
     \\[-1.8ex]\hline
     \hline \\[-1.8ex]
     & \multicolumn{6}{c}{$D_i \sim \textit{Bernoulli} \left( 0.5 \right)$} & \multicolumn{6}{c}{$D_i \sim \textit{Bernoulli} \left( \estimatedPropensityScore{X_i} \right)$} \\ \cmidrule{2-7} \cmidrule{8-13} 
     & \multicolumn{3}{c}{Low heterogeneity} & \multicolumn{3}{c}{High heterogeneity} & \multicolumn{3}{c}{Low heterogeneity} & \multicolumn{3}{c}{High heterogeneity} \\ \cmidrule{2-4} \cmidrule{5-7} \cmidrule{8-10} \cmidrule{11-13} 
     & 500 & 1,000 & 2,000 & 500 & 1,000 & 2,000 & 500 & 1,000 & 2,000 & 500 & 1,000 & 2,000 \\ 
     \addlinespace[2pt]
     \hline \\[-1.8ex] 

     \multicolumn{12}{l}{\textbf{\small Panel 1: $\overline{RMSE}$}} \\
     $AT_{\scriptstyle_{XL}}$ & 238.54 & 181.90 & 136.67 & 239.59 & 188.98 & 144.79 & 303.39 & 228.18 & 173.94 & 307.69 & 237.66 & 186.39 \\
     $AT_{\scriptstyle_{CF}}$ & 224.59 & 175.43 & 132.89 & 231.20 & 183.63 & 140.72 & 284.92 & 208.18 & 160.66 & 290.06 & 220.61 & 172.21 \\
     $CT$ & 306.66 & 304.49 & 294.14 & 318.38 & 307.98 & 302.31 & 280.94 & 273.06 & 267.65 & 283.52 & 280.03 & 270.30 \\ \cmidrule{1-13} 

     \multicolumn{12}{l}{\textbf{\small Panel 2: $\overline{\left\vert Bias \right\vert}$}} \\
     $AT_{\scriptstyle_{XL}}$ & 17.07 & 16.55 & 16.26 & 40.20 & 40.00 & 39.89 & 22.17 & 23.46 & 21.68 & 44.85 & 46.99 & 43.22 \\
     $AT_{\scriptstyle_{CF}}$ & 16.82 & 16.53 & 16.24 & 40.19 & 40.07 & 39.85 & 23.10 & 23.11 & 21.43 & 45.55 & 47.80 & 42.98 \\
     $CT$ & 17.59 & 17.63 & 17.48 & 40.76 & 40.56 & 40.47 & 18.81 & 18.21 & 18.45 & 40.45 & 41.49 & 40.49 \\ \cmidrule{1-13} 

     \multicolumn{12}{l}{\textbf{\small Panel 3: $\overline{SD}$}} \\
     $AT_{\scriptstyle_{XL}}$ & 237.56 & 180.68 & 135.07 & 234.21 & 182.17 & 135.89 & 302.08 & 226.25 & 171.78 & 302.38 & 230.22 & 178.25 \\
     $AT_{\scriptstyle_{CF}}$ & 223.59 & 174.19 & 131.30 & 225.71 & 176.74 & 131.85 & 283.42 & 206.15 & 158.38 & 284.34 & 212.40 & 163.56 \\
     $CT$ & 305.86 & 303.67 & 293.31 & 314.28 & 303.74 & 298.01 & 279.94 & 272.10 & 266.63 & 278.99 & 275.22 & 265.52 \\ \cmidrule{1-13} 

     \multicolumn{12}{l}{\textbf{\small Panel 4: Coverage of $95\%$ CI}} \\
     $AT_{\scriptstyle_{XL}}$ & 0.92 & 0.94 & 0.94 & 0.93 & 0.93 & 0.94 & 0.93 & 0.94 & 0.93 & 0.92 & 0.92 & 0.92 \\
     $AT_{\scriptstyle_{CF}}$ & 0.93 & 0.94 & 0.94 & 0.94 & 0.94 & 0.94 & 0.94 & 0.94 & 0.94 & 0.93 & 0.92 & 0.92 \\
     $CT$ & 0.89 & 0.89 & 0.89 & 0.89 & 0.89 & 0.89 & 0.85 & 0.85 & 0.85 & 0.84 & 0.85 & 0.85 \\ \cmidrule{1-13} 

     \multicolumn{12}{l}{\textbf{\small Panel 5: $\overline{|\mathcal{T}|}$}} \\
     $AT_{\scriptstyle_{XL}}$ & 9.58 & 11.75 & 13.36 & 9.46 & 11.93 & 13.43 & 8.38 & 10.39 & 11.96 & 8.32 & 10.37 & 12.07 \\
     $AT_{\scriptstyle_{CF}}$ & 8.83 & 11.26 & 12.84 & 9.04 & 11.37 & 12.96 & 7.40 & 8.96 & 10.80 & 7.51 & 8.86 & 10.68 \\
     $CT$ & 15.29 & 29.81 & 56.95 & 15.51 & 29.96 & 57.77 & 6.89 & 13.29 & 25.33 & 6.78 & 13.37 & 24.69 \\ 

     \hline
     \hline \\[-1.8ex]
  \end{tabular}
  \end{adjustbox}
  \caption{Comparison with causal trees. The first three panels report the average over the validation sample of $RMSE$ ($\overline{RMSE}$), $\left\vert Bias \right\vert$ ($\overline{\left\vert Bias \right\vert}$), and $SD$ ($\overline{SD}$). The fourth panel reports coverage rates for $95\%$ confidence intervals. The last panel reports the average number of leaves. All trees are honest.}
  \label{table_simulation_results_honest} 
  \end{table}
\endgroup

The second and third panels of Table \ref{table_simulation_results_honest} offer additional insights into the superior prediction performance of aggregation trees by examining the absolute bias and standard deviation of the estimators. As expected, the advantage of aggregation trees is entirely driven by their lower sampling variance, resulting from the more stable splitting strategy employed. All estimators exhibit some bias, which increases as heterogeneity strengthens. When treatment is randomly assigned, the bias across all estimators is approximately the same. When treatment is instead assigned based on $\estimatedPropensityScore{\cdot}$, causal trees exhibit a slightly lower bias than both $AT_{\scriptstyle_{XL}}$ and $AT_{\scriptstyle_{CF}}$. However, aggregation trees show substantially lower sampling variance than causal trees, with approximately the same magnitudes and exceptions noted for RMSE. For all estimators, sampling variance remains consistent across different heterogeneity levels.

The fourth panel of Table \ref{table_simulation_results_honest} presents the coverage rates for $95\%$ confidence intervals. Both $AT_{\scriptstyle_{XL}}$ and $AT_{\scriptstyle_{CF}}$ achieve coverage rates close to the nominal level, while causal trees consistently show lower coverage. The fifth panel of Table \ref{table_simulation_results_honest} reveals that causal trees generate more complex models with a larger number of leaves compared to aggregation trees.\footnote{\ \textcolor{purple}{This pattern is consistent with prior evidence that causal trees tend to grow larger than alternative tree methods \parencite[see, e.g.,][]{lee2021discovering, yang2022causal}.}} The complexity of causal trees increases significantly with larger sample sizes, while that of $AT_{\scriptstyle_{XL}}$ and $AT_{\scriptstyle_{CF}}$ grows more conservatively. This could lead to overfitting and raise concerns about multiple hypothesis testing in causal trees, potentially explaining their lower coverage rates. For all estimators, model complexity does not vary across heterogeneity levels and is smaller when treatment is assigned based on $\estimatedPropensityScore{\cdot}$, particularly for causal trees.

Comparing these results with Table \ref{table_simulation_results_adaptive} allows us to assess the benefits of honesty. Using different data for constructing the trees and treatment effect estimation greatly benefits inference. The coverage rates of adaptive trees are considerably below the nominal rate, particularly those of causal trees that can be as low as $58\%$. 

While honesty is expected to come at the expense of a larger mean squared error, our simulations show that the RMSE is actually higher for adaptive trees than for honest trees. For aggregation trees, this discrepancy may be due to the slightly more complex models generated by adaptive estimation, resulting in higher sampling variance while maintaining the same bias. In the case of causal trees, the discrepancy may also be due to the greater bias induced by adaptive estimation when the assignment is based on $\estimatedPropensityScore{\cdot}$. An exception arises for causal trees under randomized treatment assignment, where adaptivity produces extremely shallow trees, which results in lower RMSE compared to honest estimation.

%%%%%%%%%%%%%%%%%%%%%%%%%%%%%%%%%%%%%%%%
%        EMPIRICAL ILLUSTRATION.       %
%%%%%%%%%%%%%%%%%%%%%%%%%%%%%%%%%%%%%%%%

\section{Empirical illustration} 
\label{sec_empirical_illustration}
\noindent In this section, we apply the methodology developed in this paper to revisit the impact of maternal smoking on birth weight using the same data set as in the previous section. First, we review previous findings from the literature. Then, we construct the sequence of optimal groupings. Finally, we discuss the results.

%%% PREVIOUS FINDINGS   
\subsection{Previous findings}
\label{subsection_previous_findings}
\noindent As documented in \textcite{almond2005costs}, infants born at LBW can impose substantial costs on society, with estimated expected costs of delivery and initial care exceeding 100,000\$ (at prices of year 2000) for babies weighing 1,000 grams at birth.\footnote{\ An infant is considered born at LBW if she weighs less than 2,500 grams at birth.} Moreover, LBW is associated with a higher risk of death within one year of birth. For these reasons, birth weight is considered the primary measure of a baby’s health and is often the direct target of health policies. Thus, understanding what causes LBW is crucial.

The impact of maternal smoking on LBW has received considerable attention in the literature, for it is regarded as one of the most significant and modifiable risk factors. Several studies consistently find that smoking during pregnancy causes lower average birth weight, with estimated ATEs ranging between -600 and -100 grams \parencite[see, e.g.,][]{almond2005costs, abrevaya2006estimating}. As for effect heterogeneity, it is now well understood that the effects are increasingly negative with the mother's age \parencite{abrevaya2015estimating, lee2017doubly, zimmert2019nonparametric, fan2022estimation}. Finally, treatment heterogeneity has also been investigated. \textcite{cattaneo2010efficient} and \textcite{bodory2022high} consider different smoking intensities as different treatments and show that higher smoking intensities lead to more negative effects. \textcite{heiler2021effect} find that heterogeneous effects can be partly explained by different smoking behaviors of ethnic and age groups.

%% CONSTRUCTING THE SEQUENCE OF GROUPINGS.
\subsection{Constructing the sequence of groupings}
\label{subsec_empirical_illustration_construction_sequence_groupings}
\noindent We split the sample into a training sample and an honest sample of equal sizes. We estimate the CATEs by fitting an honest causal forest \parencite{athey2019generalized}.\footnote{\ Figure \ref{fig_sorted_cates} in Appendix \ref{app_data_description} displays the estimated CATEs sorted by magnitude, along with their corresponding $95\%$ confidence intervals. Almost all predicted effects are negative and statistically different from zero at the $5\%$ significance level.} We then construct the set of admissible groupings by approximating the estimated CATEs with a decision tree.\footnote{\ \textcolor{purple}{Tree construction uses the R package \texttt{rpart} with default settings, except for the complexity parameter \texttt{cp}---which controls the minimum reduction in cost–complexity risk required for a split---which we set to $0.02$ (the default is $0.01$) to obtain a shallower initial tree and improve plot readability. As an alternative readability constraint, one may limit the maximum depth of the tree \parencite[see, e.g.,][]{lee2021discovering, bargagli2022heterogeneous}.}} We perform these operations utilizing only observations from the training sample.

Once the tree is constructed, we use observations from the honest sample to estimate the GATEs by constructing and averaging the doubly-robust scores in (\ref{equation_dr_scores}). Honest regression forests and $5$-fold cross-fitting are employed to estimate the necessary nuisance functions.

\begin{figure}[b!]
    \centering
    \includegraphics[scale=0.5]{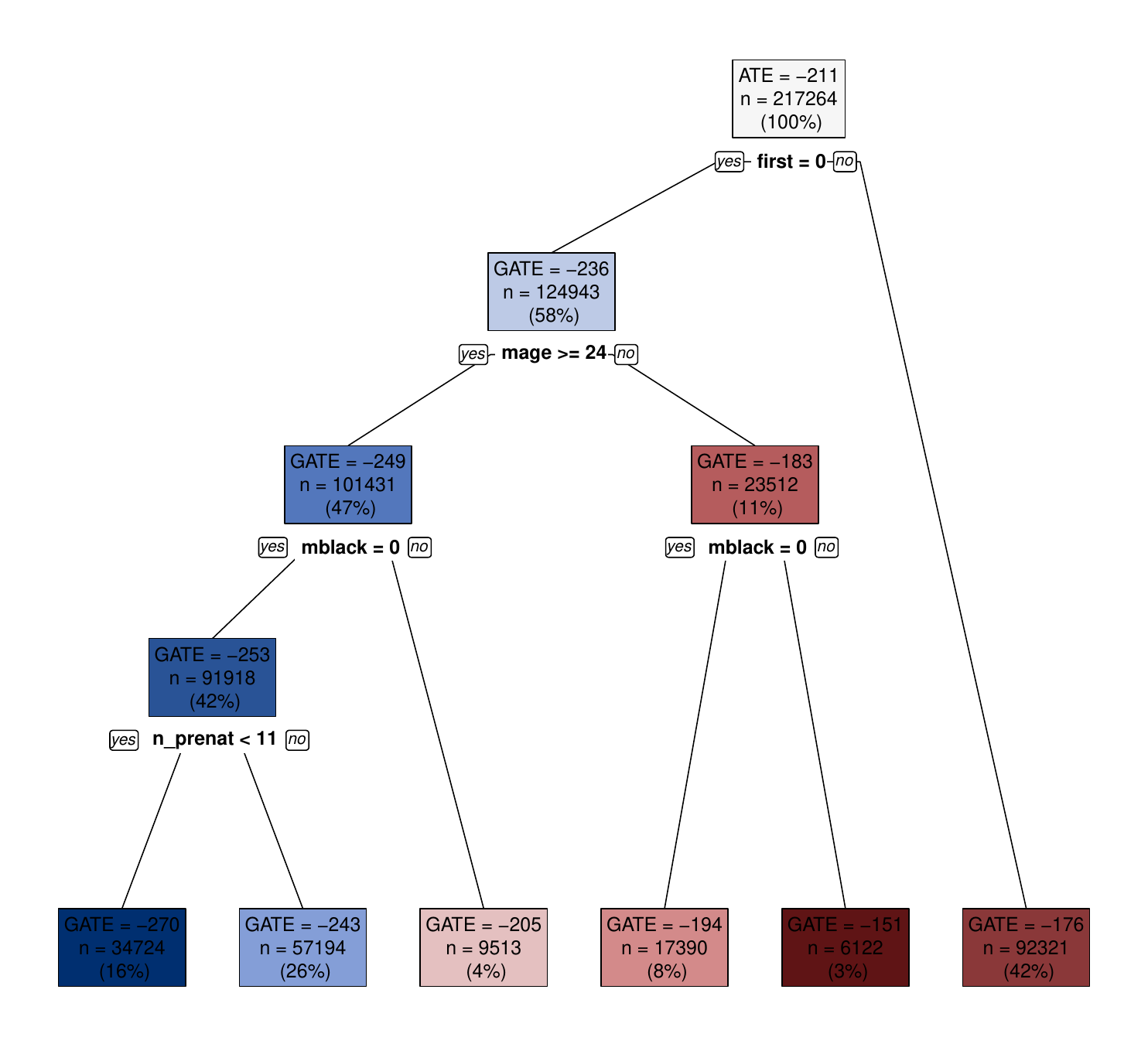}
    \caption{Aggregation tree, constructed in the training sample. Each node displays the GATE and the number and percentage of units belonging to each subgroup. The GATEs are estimated by averaging doubly-robust scores constructed using the honest sample. Blue and red shades denote groups with GATEs stronger (i.e., more negative) and lighter (i.e., more positive) than the ATE.}
    \label{fig_aggregation_tree}
\end{figure}

Figure \ref{fig_aggregation_tree} displays the resulting tree. At the root, we see the estimated ATE of $-211$ grams. \textcolor{purple}{As detailed in Section \ref{subsec_treegrowing}, the algorithm then chooses the splitting variable and point that most reduce the within-leaf variation of the predicted CATEs---i.e., the split that maximizes between-group heterogeneity---and repeats this recursively in the child nodes.} The observations are initially split into two groups: non-first-born children (root's left child) and first-born children (root's right child). This division, among all possible two-group splits, maximizes heterogeneity in treatment effects. The first group, which includes $58\%$ of the units, has an estimated GATE of approximately $-236$ grams, while the second group, representing $42\%$ of the units, has an estimated GATE of about $-176$ grams.

\textcolor{purple}{The non–first-born branch is further split by mother’s age (younger vs. older), yielding two new groups: non–first-born children with older mothers ($47\%$ of the sample; estimated GATE $\approx-249$) and with younger mothers ($11\%$; estimated GATE $\approx-183$). Both groups are further partitioned by maternal race (Black vs. non-Black), and one of the resulting nodes is split again by the number of prenatal care visits. In total, this produces six terminal groups.}

\begin{figure}[t!]
    \centering
    \makebox[\textwidth][c]{
    \subfloat{\includegraphics[scale=0.32]{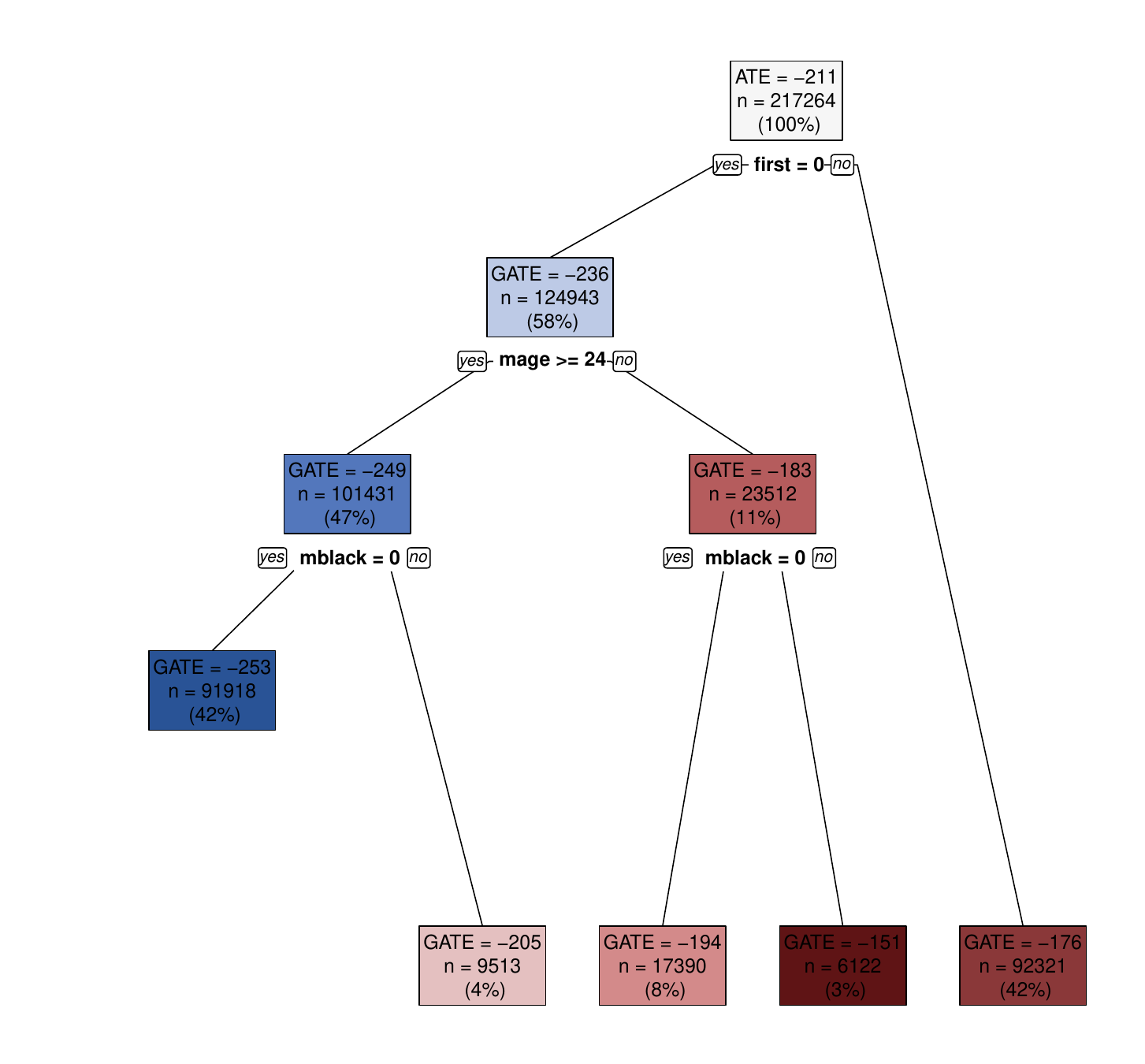}}
    \subfloat{\includegraphics[scale=0.32]{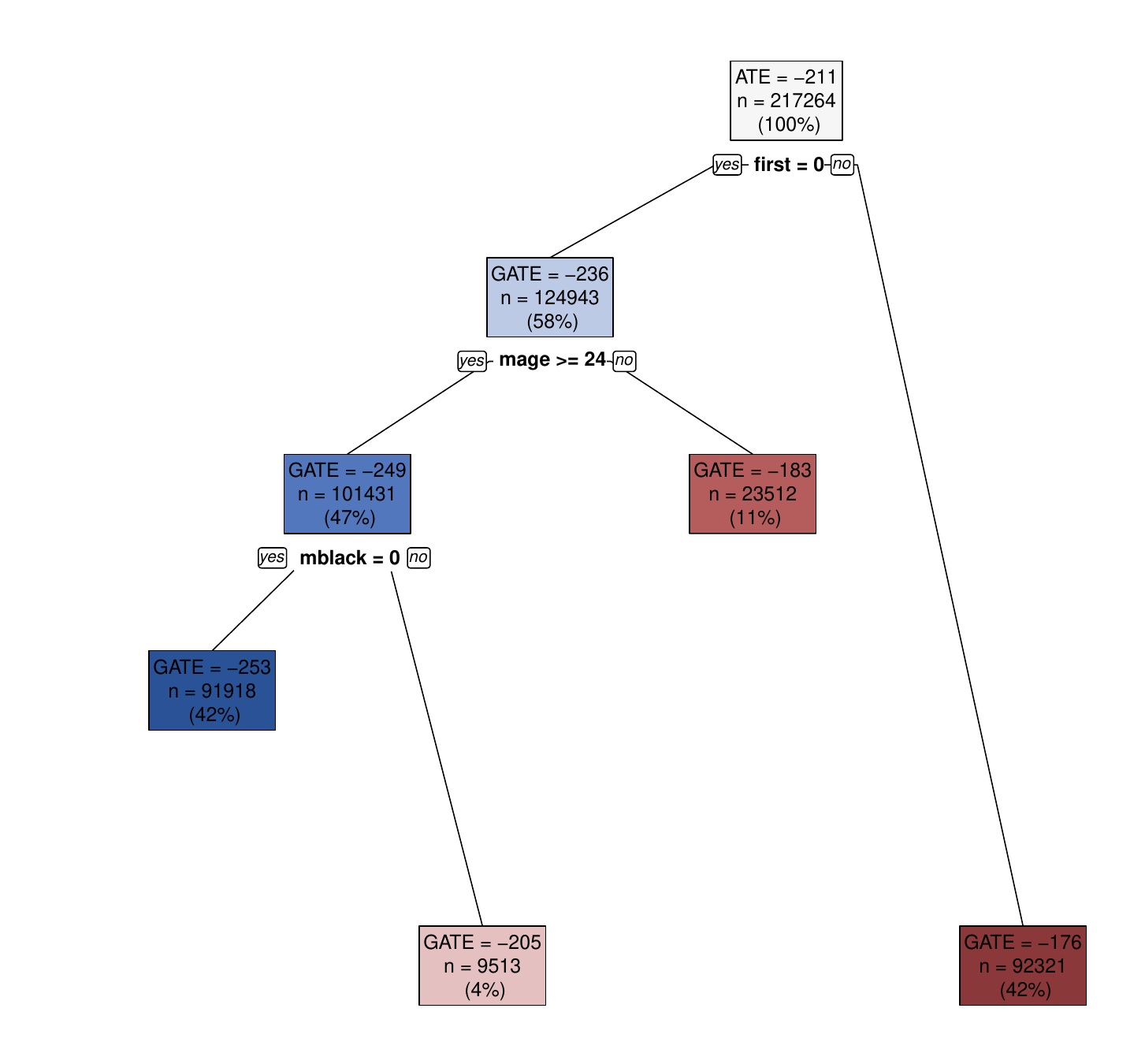}}}\\
    \makebox[\textwidth][c]{\subfloat{\includegraphics[scale=0.32]{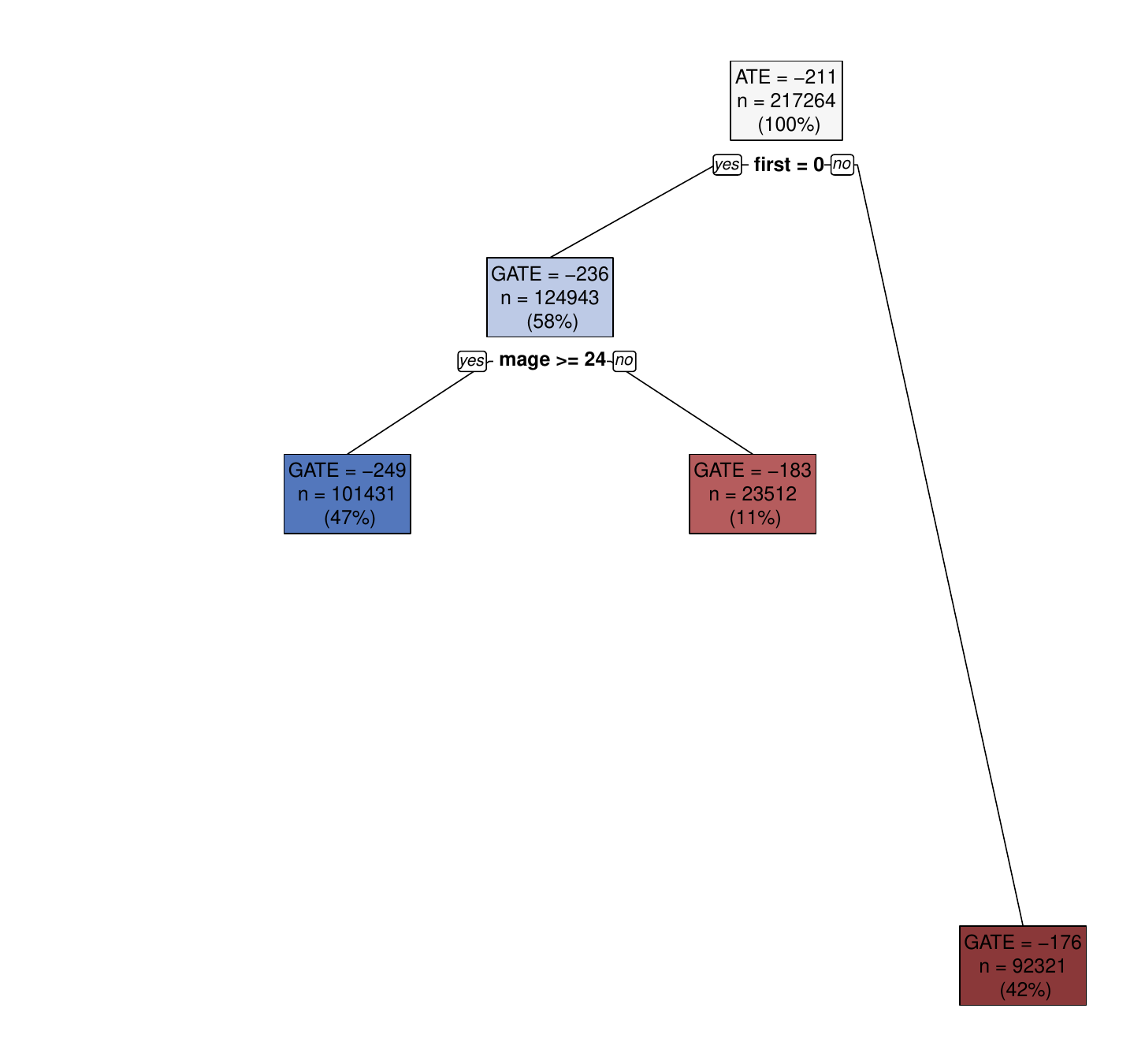}}
    \subfloat{\includegraphics[scale=0.32]{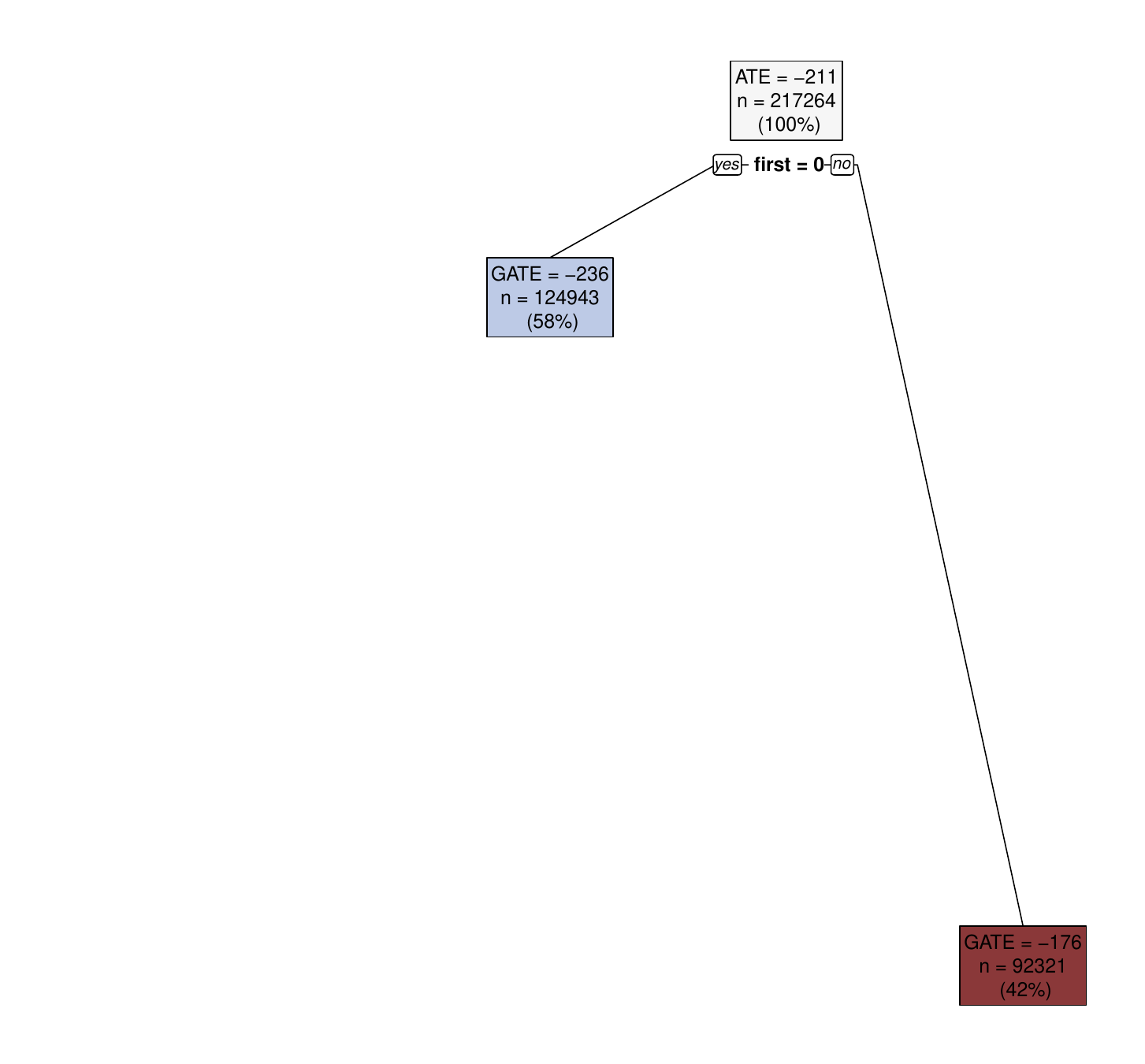}}}%
    \caption{Sequence of optimal groupings, obtained by progressively collapsing the node for which the loss in explained heterogeneity is minimized.}
    \label{fig_sequence_groupings}
\end{figure}

The sequence of optimal groupings is then constructed by progressively aggregating the two subgroups that yield the smallest loss in explained heterogeneity. Figure \ref{fig_sequence_groupings} visually illustrates this procedure.\footnote{\ \textcolor{purple}{Figure~\ref{fig_alpha_path_dual} in Appendix \ref{app_data_description} reports cross-validated risk along the pruning path as a function of the number of leaves and of the cost–complexity parameter $\alpha$.}} Reading the figure from left to right and top to bottom, each panel corresponds to a grouping in the sequence, derived by collapsing the node that minimizes the loss in explained heterogeneity. The figure clearly demonstrates the consistency of the results across the different granularity levels, highlighting how the generated sequence enables a coherent exploration of the trade-off between interpretability and the discovery of more granular heterogeneity.

%% DISCUSSION.
\subsection{Discussion}
\label{subsec_empirical_discussion}

\begingroup
  \setlength{\tabcolsep}{8pt}
  \renewcommand{\arraystretch}{1.2}
  \begin{table}[b!]
    \centering
    \begin{adjustbox}{width = 1\textwidth}
    \begin{tabular}{@{\extracolsep{5pt}}l c c c c c}
      \\[-1.8ex]\hline
      \hline \\[-1.8ex] 

      & \textit{Leaf 1} & \textit{Leaf 2} & \textit{Leaf 3} & \textit{Leaf 4} & \textit{Leaf 5} \\
      \addlinespace[2pt]
      \hline \\[-1.8ex] 

      \multirow{3}{*}{GATEs} & -252.941 & -205.349 & -193.974 & -176.181 & -151.200 \\
      & [-265.761, -240.121] & [-237.959, -172.739] & [-212.363, -175.585] & [-189.003, -163.359] & [-182.001, -120.399] \\

      \addlinespace[2pt]
      \hline \\[-1.8ex] 

      \textit{Leaf 1} & - & - & - & - & - \\
            & (-) & (-) & (-) & (-) & (-) \\ 
      \textit{Leaf 2} & 47.59 &    - &    - &    - & - \\
            & (0.054) & (   -) & (   -) & (   -) & (-) \\ 
      \textit{Leaf 3} & 58.97 & 11.38 &    - &    - & - \\
            & (0.000) & (0.551) & (   -) & (   -) & (-) \\ 
      \textit{Leaf 4} & 76.76 & 29.17 & 17.79 &    - & - \\
            & (0.000) & (0.411) & (0.411) & (   -) & (-) \\ 
      \textit{Leaf 5} & 101.74 &  54.15 &  42.77 &  24.98 & - \\
            & (0.000) & (0.108) & (0.108) & (0.411) & (-) \\ 

      \addlinespace[3pt]
      \\[-1.8ex]\hline
      \hline \\[-1.8ex]
    \end{tabular}
    \end{adjustbox}
    \caption{Point estimates and $95\%$ confidence intervals for the GATEs. Leaves are sorted in increasing order of the GATEs. Additionally, the GATE differences across all pairs of leaves are displayed. $p$-values testing the null hypothesis that a single difference is zero are adjusted using Holm's procedure and reported in parenthesis under each point estimate.}
    \label{table_differences_gates}
    \end{table}
\endgroup 

\noindent We investigate whether systematic effect heterogeneity is present by examining whether distinct subgroups exhibit different reactions to the treatment.\footnote{\ Looking at the distribution of the estimated CATEs is not an effective strategy for this task, as high variation in predictions due to estimation noise does not necessarily imply heterogeneous effects.} For this purpose, we select the optimal grouping composed of five groups and estimate model (\ref{equation_observational_studies_ols}) using only observations from the honest sample. Table \ref{table_differences_gates} reports point estimates and $95\%$ confidence intervals.\footnote{\ \textcolor{purple}{We report five groups to provide a practical summary that balances interpretability and detail. Results are robust to adjacent granularities: Tables \ref{table_differences_gates_four} and \ref{table_differences_gates_six} in Appendix \ref{app_data_description} show partitions with four and six groups, respectively, yielding similar GATE patterns and conclusions. This robustness is expected because the tree-pruning step produces a nested sequence of partitions (see Section \ref{subsec_treepruning}), thus implying that nearby values of $\alpha$ generate closely related groupings and similar GATE patterns.}} The estimated GATEs exhibit substantial differences, ranging from $-252$ grams for the most affected group (\textit{Leaf 1}) to $-151$ grams for the least affected group (\textit{Leaf 5}). All estimates are negative and statistically different from zero at the $5\%$ significance level. 

Table \ref{table_differences_gates} also reports the GATE differences across all pairs of groups, along with $p$-values testing the null hypothesis that each difference equals zero. To account for multiple hypothesis testing, we adjust the $p$-values using the procedure of \textcite{holm1979multiple}. Many groups exhibit significant differences. For example, the GATE differences between \textit{Leaf 1} and all other groups are both large and statistically significant (at the $10\%$ confidence level for \textit{Leaf 2} and $5\%$ for the other groups). Additionally, the $p$-values for the differences between \textit{Leaf 5} and \textit{Leaf 2} or \textit{Leaf 3} are just above the $10\%$ significance level. For other pairs, GATE differences are small, and we fail to reject the null hypothesis at any conventional confidence level. Overall, these findings provide evidence of systematic heterogeneity in treatment effects.

\begingroup
  \setlength{\tabcolsep}{8pt}
  \renewcommand{\arraystretch}{1.1}
  \begin{table}[b!]
    \centering
    \begin{adjustbox}{width = 1\textwidth}
    \begin{tabular}{@{\extracolsep{5pt}}l c c c c c c c c c c c}
      \\[-1.8ex]\hline
      \hline \\[-1.8ex]
      & \multicolumn{2}{c}{\textit{Leaf 1}} & \multicolumn{2}{c}{\textit{Leaf 2}} & \multicolumn{2}{c}{\textit{Leaf 3}} & \multicolumn{2}{c}{\textit{Leaf 4}} & \multicolumn{2}{c}{\textit{Leaf 5}} \\\cmidrule{2-3} \cmidrule{4-5} \cmidrule{6-7} \cmidrule{8-9} \cmidrule{10-11} 
      & Mean & (S.E.) & Mean & (S.E.) & Mean & (S.E.) & Mean & (S.E.) & Mean & (S.E.) \\
      \addlinespace[2pt]
      \hline \\[-1.8ex] 

      \multicolumn{9}{l}{\textbf{Parental characteristics}} \\
      \texttt{mage} & 30.296 & (0.013) & 29.275 & (0.043) & 21.185 & (0.013) & 25.009 & (0.018) & 20.408 & (0.025) \\ 
      \texttt{meduc} & 13.171 & (0.008) & 12.442 & (0.024) & 11.229 & (0.013) & 12.972 & (0.008) & 11.188 & (0.027) \\ 
      \texttt{fage} & 32.334 & (0.017) & 31.914 & (0.069) & 24.780 & (0.033) & 27.512 & (0.021) & 23.546 & (0.058) \\ 
      \texttt{feduc} & 13.294 & (0.009) & 11.884 & (0.037) & 11.338 & (0.018) & 12.886 & (0.009) & 11.003 & (0.043) \\ 

      \multicolumn{9}{l}{\textbf{Birth characteristics}} \\
      \texttt{n\_prenatal} & 11.257 & (0.010) &  8.844 & (0.046) &  9.997 & (0.029) & 11.314 & (0.011) &  6.916 & (0.057) \\ 
      \texttt{prenatal0} & 0.004 & (0.000) & 0.062 & (0.002) & 0.015 & (0.001) & 0.006 & (0.000) & 0.098 & (0.004) \\ 
      \texttt{prenatal1} & 0.870 & (0.001) & 0.614 & (0.005) & 0.677 & (0.004) & 0.836 & (0.001) & 0.405 & (0.006) \\ 
      \texttt{prenatal2} & 0.103 & (0.001) & 0.243 & (0.004) & 0.243 & (0.003) & 0.131 & (0.001) & 0.337 & (0.006) \\ 
      \texttt{prenatal3} & 0.019 & (0.000) & 0.071 & (0.003) & 0.060 & (0.002) & 0.023 & (0.000) & 0.147 & (0.005) \\ 

      \addlinespace[3pt]
      \\[-1.8ex]\hline
      \hline \\[-1.8ex]
    \end{tabular}
    \end{adjustbox}
    \caption{Average characteristics of units in each leaf, obtained by regressing each covariate on a set of dummies denoting leaf membership using only the honest sample. Standard errors are estimated via the Eicker-Huber-White estimator. Leaves are sorted in increasing order of the GATEs.}
    \label{table_average_characteristics_leaves}
    \end{table}
\endgroup 

To understand the factors driving this heterogeneity, we examine how treatment effects relate to observable covariates by analyzing how the average characteristics of units vary across subgroups \parencite[see, e.g.,][]{chernozhukov2017generic}.\footnote{\ Another approach is to assess which variables the tree-growing process used to construct groups and measure their relative importance. However, we should not conclude that covariates not used for splitting are unrelated to heterogeneity, because if two covariates are highly correlated, trees generally split on only one of them.} Table \ref{table_average_characteristics_leaves} reports the average values of selected covariates across \textit{Leaves 1}–\textit{5} (see Table \ref{table_average_characteristics_leaves_full} for the remaining covariates). The least affected group comprises children born to younger parents, suggesting more negative effects at higher parental ages. This finding is consistent with previous research \parencite[see, e.g.,][]{abrevaya2015estimating, zimmert2019nonparametric}. In contrast, the most affected group has higher parental educational attainment, likely due to older parents having had more time to study. This group also includes parents who attended more prenatal visits and were more likely to have their first visit in the first trimester of pregnancy. This may reflect mothers with problematic pregnancies self-selecting into more frequent and earlier prenatal visits.

%%%%%%%%%%%%%%%%%%%%%%%%%%%%%%%%%%%%%%%%
%              CONCLUSION.             %
%%%%%%%%%%%%%%%%%%%%%%%%%%%%%%%%%%%%%%%%

\section{Conclusion} 
\label{sec_conclusion}
\noindent This paper introduces a methodology for constructing heterogeneous subgroups that enables a flexible and coherent exploration of the trade-off between interpretability and the discovery of more granular heterogeneity. The proposed methodology constructs a sequence of groupings, one for each level of granularity. We show that each grouping features an optimality property and that the sequence is nested. We also show how the proposed methodology can be combined with honesty \parencite{athey2016recursive} and debiased machine learning procedures \parencite{semenova2021debiased} to conduct valid inference about the GATEs. 

We compare the performance of aggregation and causal trees \parencite{athey2016recursive} using both theoretical arguments and an empirical Monte-Carlo study \parencite{lechner2013sensitivity, huber2013performance}. Our simulation shows that aggregation trees substantially improve the root mean squared error of treatment effects due to lower variance resulting from a more robust splitting strategy.

We apply the proposed methodology to revisit the impact of maternal smoking on birth weight \parencite[see, e.g.,][]{almond2005costs, cattaneo2010efficient}. The analysis finds evidence of systematic heterogeneity driven by parental and birth-related characteristics.

\section*{Acknowledgements}
\noindent I especially would like to thank Franco Peracchi for feedback and suggestions. I am also grateful to Hannah Busshoff, Elena Dal Torrione, Matteo Iacopini, Michael Knaus, Michael Lechner, Giovanni Mellace, Seetha Menon, Tommaso Proietti, seminar participants at University of Rome Tor Vergata and SEW-HSG research seminars, and conference participants at the WEEE 2022, the 1st Rome Ph.D. in Economics and Finance Conference, and the ICEEE 2023 for comments and discussions. Matias Cattaneo generously provided the data used in the empirical illustration of this paper. 

The R package for implementing the methodology developed in this paper is available on CRAN at \href{https://CRAN.R-project.org/package=aggTrees}{https://CRAN.R-project.org/package=aggTrees}. The associated vignette is at \href{https://riccardo-df.github.io/aggTrees/}{https://riccardo-df.github.io/aggTrees/}.

\section*{Declaration of Interest Statement}
\noindent The author reports there are no competing interests to declare.

\section*{Data Availability Statement}
\noindent The data that support the findings of this study are subject to third-party restrictions. They were obtained under license from a scholar who does not permit public dissemination. Researchers interested in replicating or extending this work may contact the corresponding author. Access to the data will require permission from the third party that provided them.

\newpage

%%%%%%%%%%%%%%%%%%%%%%%%%%%%%%%%%%%%%%%%
%             REFERENCES.              %
%%%%%%%%%%%%%%%%%%%%%%%%%%%%%%%%%%%%%%%%

\singlespacing
\newrefcontext[sorting = nty]
\newpage
\printbibliography

%%%%%%%%%%%%%%%%%%%%%%%%%%%%%%%%%%%%%%%%
%              APPENDIX.               %
%%%%%%%%%%%%%%%%%%%%%%%%%%%%%%%%%%%%%%%%

\renewcommand\theequation{\Alph{section}.\arabic{equation}}
\renewcommand\thetable{\Alph{section}.\Roman{table}}
\renewcommand\thefigure{\Alph{section}.\Roman{figure}}

\begin{appendices}

\doublespacing

%%% DATA DESCRIPTION.
\section{Data}
\label{app_data_description}

\setcounter{equation}{0}
\setcounter{table}{0}
\setcounter{figure}{0}

\begin{table}[H]
    \setlength{\tabcolsep}{40pt}
    \centering
    \footnotesize
    \begin{adjustbox}{width = 0.92\textwidth}
        \begin{tabular}{l l}
            \midrule\midrule 
            \textbf{Label} & \textbf{Description} \\
            \midrule 
            \multicolumn{2}{l}{\textbf{OUTCOME.}} \\
            \texttt{bweight} & Infant birth weight (in grams) \\
        
            \multicolumn{2}{l}{\textbf{TREATMENT.}} \\
            \texttt{smoke} & =1 if mother smoked during pregnancy \\

            \multicolumn{2}{l}{\textbf{COVARIATES.}} \\
            \multicolumn{2}{l}{\textbf{Mother's characteristics.}} \\
            \texttt{mage} & Mother's age \\
            \texttt{meduc} & Mother's educational attainment \\
            \texttt{mwhite} & =1 if mother is white \\
            \texttt{mblack} & =1 if mother is black \\
            \texttt{mhispan} & =1 if mother is hispanic \\
        
            \texttt{foreign\_born} & =1 if mother is foreign born \\
            \texttt{unmarried} & =1 if mother is unmarried \\
            \texttt{alcohol} & =1 if mother drank alcohol during pregnancy \\
            \texttt{n\_drink} & Number of drinks per week during pregnancy \\
        
            \multicolumn{2}{l}{\textbf{Father's characteristics.}} \\
            \texttt{fage} & Father's age \\
            \texttt{feduc} & Father's educational attainment \\
            \texttt{fwhite} & =1 if father is white \\
            \texttt{fblack} & =1 if father is black \\
            \texttt{fhispan} & =1 if father is hispanic \\
        
            \multicolumn{2}{l}{\textbf{Birth characteristics.}} \\
            \texttt{birthmonth1} & =1 if birth in January \\
            \texttt{birthmonth2} & =1 if birth in February \\
            \texttt{birthmonth3} & =1 if birth in March \\
            \texttt{birthmonth4} & =1 if birth in April \\
            \texttt{birthmonth5} & =1 if birth in May \\
            \texttt{birthmonth6} & =1 if birth in June \\
            \texttt{birthmonth7} & =1 if birth in July \\
            \texttt{birthmonth8} & =1 if birth in August \\
            \texttt{birthmonth9} & =1 if birth in September \\
            \texttt{birthmonth10} & =1 if birth in October \\
            \texttt{birthmonth11} & =1 if birth in November \\
            \texttt{birthmonth12} & =1 if birth in December \\
            \texttt{first} & =1 if the infant is first born \\
            \texttt{plural} & =1 if twins or greater birth \\
            \texttt{n\_prenatal} & Number of prenatal care visits \\
            \texttt{prenatal0} & =1 if no prenatal visit \\
            \texttt{prenatal1} & =1 if first prenatal visit in first trimester of pregnancy \\
            \texttt{prenatal2} & =1 if first prenatal visit in second trimester of pregnancy \\
            \texttt{prenatal3} & =1 if first prenatal visit in third trimester of pregnancy \\
            \texttt{adequacy1} & =1 if adequacy of care is adequate (Kessner Index) \\
            \texttt{adequacy2} & =1 if adequacy of care is intermediate (Kessner Index) \\
            \texttt{adequacy3} & =1 if adequacy of care is inadequate (Kessner Index) \\
        
            \multicolumn{2}{l}{\textbf{Maternal medical risk factors.}} \\ 
            \texttt{diabete} & =1 if mother is diabetic \\
            \texttt{anemia} & =1 if mother is anemic \\
            \texttt{hyper} & =1 if mother had pregnancy-associated hypertension \\
        
            \midrule 
        \end{tabular}
        \end{adjustbox}
    \caption{Description of variables in the data set.}
    \label{table_data_description}
\end{table}

\begingroup
  \setlength{\tabcolsep}{8pt}
  \renewcommand{\arraystretch}{1.1}
  \begin{table}[H]
    \centering
    \begin{adjustbox}{width = 0.75\textwidth}
    \begin{tabular}{@{\extracolsep{5pt}}l c c c c c c}
    \\[-1.8ex]\hline
    \hline \\[-1.8ex]
    & \multicolumn{2}{c}{Treated} & \multicolumn{2}{c}{Controls} & \multicolumn{2}{c}{Overlap measures} \\ \cmidrule{6-7}
    & \multicolumn{2}{c}{($n_t =  81,388 $)} & \multicolumn{2}{c}{($n_c = 353,140 $)} & \\ \cmidrule{2-5}
    & Mean & (S.D.) & Mean & (S.D.) & $\hat{\Delta}_j$ & $\hat{\Gamma}_j$ \\
    \addlinespace[2pt]
    \hline \\[-1.8ex] 

    \texttt{mage} & 25.503 & (5.372) & 27.340 & (5.553) & -0.336 & -0.033 \\ 
    \texttt{meduc} & 11.783 & (1.883) & 13.088 & (2.430) & -0.600 & -0.255 \\ 
    \texttt{mwhite} & 0.850 & (0.357) & 0.865 & (0.342) & -0.043 & 0.044 \\ 
    \texttt{mblack} & 0.147 & (0.354) & 0.116 & (0.321) & 0.090 & 0.099 \\ 
    \texttt{mhispan} & 0.020 & (0.140) & 0.031 & (0.173) & -0.068 & -0.209 \\ 
    \texttt{foreign\_born} & 0.019 & (0.138) & 0.056 & (0.229) & -0.190 & -0.504 \\ 
    \texttt{unmarried} & 0.444 & (0.497) & 0.196 & (0.397) & 0.552 & 0.225 \\ 
    \texttt{alcohol} & 0.045 & (0.207) & 0.007 & (0.080) & 0.243 & 0.943 \\ 
    \texttt{n\_drink} & 0.123 & (0.729) & 0.013 & (0.212) & 0.205 & 1.237 \\ 
    \texttt{fage} & 28.451 & (6.556) & 29.640 & (6.264) & -0.185 & 0.046 \\ 
    \texttt{feduc} & 11.686 & (2.628) & 13.102 & (2.800) & -0.522 & -0.064 \\ 
    \texttt{fwhite} & 0.831 & (0.375) & 0.857 & (0.350) & -0.072 & 0.068 \\ 
    \texttt{fblack} & 0.162 & (0.369) & 0.123 & (0.329) & 0.112 & 0.115 \\ 
    \texttt{fhispan} & 0.028 & (0.166) & 0.033 & (0.178) & -0.025 & -0.068 \\ 
    \texttt{birthmonth1} & 0.081 & (0.273) & 0.078 & (0.268) & 0.011 & 0.017 \\ 
    \texttt{birthmonth2} & 0.074 & (0.262) & 0.075 & (0.264) & -0.003 & -0.004 \\ 
    \texttt{birthmonth3} & 0.082 & (0.274) & 0.086 & (0.280) & -0.015 & -0.023 \\ 
    \texttt{birthmonth4} & 0.076 & (0.265) & 0.083 & (0.277) & -0.027 & -0.043 \\ 
    \texttt{birthmonth5} & 0.081 & (0.273) & 0.087 & (0.282) & -0.022 & -0.032 \\ 
    \texttt{birthmonth6} & 0.083 & (0.277) & 0.086 & (0.280) & -0.009 & -0.013 \\ 
    \texttt{birthmonth7} & 0.092 & (0.289) & 0.089 & (0.284) & 0.011 & 0.016 \\ 
    \texttt{birthmonth8} & 0.094 & (0.291) & 0.089 & (0.285) & 0.017 & 0.024 \\ 
    \texttt{birthmonth9} & 0.090 & (0.286) & 0.087 & (0.282) & 0.011 & 0.016 \\ 
    \texttt{birthmonth10} & 0.086 & (0.281) & 0.084 & (0.277) & 0.009 & 0.013 \\ 
    \texttt{birthmonth11} & 0.078 & (0.269) & 0.077 & (0.267) & 0.003 & 0.005 \\ 
    \texttt{birthmonth12} & 0.082 & (0.275) & 0.079 & (0.270) & 0.012 & 0.019 \\ 
    \texttt{first} & 0.367 & (0.482) & 0.438 & (0.496) & -0.146 & -0.029 \\ 
    \texttt{plural} & 0.015 & (0.120) & 0.016 & (0.127) & -0.015 & -0.060 \\ 
    \texttt{n\_prenatal} & 10.210 & (3.989) & 11.125 & (3.395) & -0.247 & 0.161 \\ 
    \texttt{prenatal0} & 0.025 & (0.156) & 0.007 & (0.086) & 0.141 & 0.603 \\ 
    \texttt{prenatal1} & 0.718 & (0.450) & 0.838 & (0.368) & -0.292 & 0.200 \\ 
    \texttt{prenatal2} & 0.204 & (0.403) & 0.124 & (0.330) & 0.216 & 0.200 \\ 
    \texttt{prenatal3} & 0.047 & (0.212) & 0.026 & (0.159) & 0.114 & 0.289 \\ 
    \texttt{adequacy1} & 0.631 & (0.483) & 0.762 & (0.426) & -0.287 & 0.125 \\ 
    \texttt{adequacy2} & 0.258 & (0.437) & 0.184 & (0.388) & 0.178 & 0.121 \\ 
    \texttt{adequacy3} & 0.105 & (0.306) & 0.049 & (0.216) & 0.210 & 0.347 \\ 
    \texttt{diabete} & 0.018 & (0.132) & 0.018 & (0.135) & -0.006 & -0.022 \\ 
    \texttt{anemia} & 0.014 & (0.119) & 0.008 & (0.092) & 0.057 & 0.266 \\ 
    \texttt{hyper} & 0.019 & (0.138) & 0.029 & (0.168) & -0.065 & -0.202 \\ 

    \addlinespace[3pt]
    \\[-1.8ex]\hline
    \hline \\[-1.8ex]
    \end{tabular}
    \end{adjustbox}
    \caption{Balance between treatment and control groups. The last two columns report the estimated normalized differences ($\hat{\Delta}_j$) and logarithms of the ratio of standard deviations ($\hat{\Gamma}_j$).}
    \label{table_descriptive_stats}
    \end{table}
\endgroup

\begingroup
  \setlength{\tabcolsep}{8pt}
  \renewcommand{\arraystretch}{1.1}
  \begin{table}[H]
    \centering
    \begin{adjustbox}{width = 1\textwidth}
    \begin{tabular}{@{\extracolsep{5pt}}l c c c c c c c c c c}
      \\[-1.8ex]\hline
      \hline \\[-1.8ex]
      & \multicolumn{2}{c}{\textit{Leaf 1}} & \multicolumn{2}{c}{\textit{Leaf 2}} & \multicolumn{2}{c}{\textit{Leaf 3}} & \multicolumn{2}{c}{\textit{Leaf 4}} & \multicolumn{2}{c}{\textit{Leaf 5}} \\\cmidrule{2-3} \cmidrule{4-5} \cmidrule{6-7} \cmidrule{8-9} \cmidrule{10-11} 
      & Mean & (S.E.) & Mean & (S.E.) & Mean & (S.E.) & Mean & (S.E.) & Mean & (S.E.) \\
      \addlinespace[2pt]
      \hline \\[-1.8ex] 

      \texttt{mwhite} & 0.983 & (0.000) & 0.000 & (-) & 0.988 & (0.001) & 0.863 & (0.001) & 0.000 & (-) \\ 
      \texttt{mblack} & 0.000 & (-) & 1.000 & (-) & 0.000 & (-) & 0.118 & (0.001) & 1.000 & (-) \\ 
      \texttt{mhispan} & 0.022 & (0.000) & 0.010 & (0.001) & 0.093 & (0.002) & 0.026 & (0.001) & 0.013 & (0.001) \\ 
      \texttt{foreign\_born} & 0.051 & (0.001) & 0.050 & (0.002) & 0.059 & (0.002) & 0.048 & (0.001) & 0.014 & (0.002) \\ 
      \texttt{unmarried} & 0.086 & (0.001) & 0.599 & (0.005) & 0.356 & (0.004) & 0.299 & (0.002) & 0.867 & (0.004) \\ 
      \texttt{alcohol} & 0.013 & (0.000) & 0.049 & (0.002) & 0.011 & (0.001) & 0.010 & (0.000) & 0.034 & (0.002) \\ 
      \texttt{n\_drink} & 0.029 & (0.001) & 0.168 & (0.010) & 0.026 & (0.002) & 0.022 & (0.001) & 0.106 & (0.009) \\ 
      \texttt{fwhite} & 0.972 & (0.001) & 0.028 & (0.002) & 0.948 & (0.002) & 0.853 & (0.001) & 0.029 & (0.002) \\ 
      \texttt{fblack} & 0.010 & (0.000) & 0.962 & (0.002) & 0.037 & (0.001) & 0.128 & (0.001) & 0.963 & (0.002) \\ 
      \texttt{fhispan} & 0.024 & (0.001) & 0.020 & (0.001) & 0.097 & (0.002) & 0.029 & (0.001) & 0.028 & (0.002) \\ 
      \texttt{birthmonth1} & 0.078 & (0.001) & 0.081 & (0.003) & 0.079 & (0.002) & 0.078 & (0.001) & 0.085 & (0.004) \\ 
      \texttt{birthmonth2} & 0.075 & (0.001) & 0.074 & (0.003) & 0.074 & (0.002) & 0.075 & (0.001) & 0.080 & (0.003) \\ 
      \texttt{birthmonth3} & 0.086 & (0.001) & 0.082 & (0.003) & 0.081 & (0.002) & 0.085 & (0.001) & 0.084 & (0.004) \\ 
      \texttt{birthmonth4} & 0.083 & (0.001) & 0.079 & (0.003) & 0.078 & (0.002) & 0.083 & (0.001) & 0.076 & (0.003) \\ 
      \texttt{birthmonth5} & 0.090 & (0.001) & 0.077 & (0.003) & 0.084 & (0.002) & 0.084 & (0.001) & 0.081 & (0.003) \\ 
      \texttt{birthmonth6} & 0.087 & (0.001) & 0.082 & (0.003) & 0.088 & (0.002) & 0.085 & (0.001) & 0.090 & (0.004) \\ 
      \texttt{birthmonth7} & 0.087 & (0.001) & 0.096 & (0.003) & 0.095 & (0.002) & 0.088 & (0.001) & 0.091 & (0.004) \\ 
      \texttt{birthmonth8} & 0.089 & (0.001) & 0.091 & (0.003) & 0.094 & (0.002) & 0.089 & (0.001) & 0.088 & (0.004) \\ 
      \texttt{birthmonth9} & 0.086 & (0.001) & 0.087 & (0.003) & 0.086 & (0.002) & 0.089 & (0.001) & 0.079 & (0.003) \\ 
      \texttt{birthmonth10} & 0.085 & (0.001) & 0.084 & (0.003) & 0.082 & (0.002) & 0.085 & (0.001) & 0.078 & (0.003) \\ 
      \texttt{birthmonth11} & 0.077 & (0.001) & 0.083 & (0.003) & 0.079 & (0.002) & 0.079 & (0.001) & 0.082 & (0.004) \\ 
      \texttt{birthmonth12} & 0.078 & (0.001) & 0.085 & (0.003) & 0.080 & (0.002) & 0.081 & (0.001) & 0.086 & (0.004) \\ 
      \texttt{first} & 0.000 & (-) & 0.000 & (-) & 0.000 & (-) & 1.000 & (-) & 0.000 & (-) \\ 
      \texttt{plural} & 0.016 & (0.000) & 0.021 & (0.001) & 0.012 & (0.001) & 0.016 & (0.000) & 0.017 & (0.002) \\ 
      \texttt{adequacy1} & 0.794 & (0.001) & 0.500 & (0.005) & 0.578 & (0.004) & 0.766 & (0.001) & 0.280 & (0.006) \\ 
      \texttt{adequacy2} & 0.165 & (0.001) & 0.295 & (0.005) & 0.306 & (0.003) & 0.189 & (0.001) & 0.356 & (0.006) \\ 
      \texttt{adequacy3} & 0.037 & (0.001) & 0.194 & (0.004) & 0.110 & (0.002) & 0.040 & (0.001) & 0.349 & (0.006) \\ 
      \texttt{diabete} & 0.021 & (0.000) & 0.026 & (0.002) & 0.009 & (0.001) & 0.017 & (0.000) & 0.007 & (0.001) \\ 
      \texttt{anemia} & 0.007 & (0.000) & 0.019 & (0.001) & 0.017 & (0.001) & 0.008 & (0.000) & 0.036 & (0.002) \\ 
      \texttt{hyper} & 0.016 & (0.000) & 0.024 & (0.002) & 0.011 & (0.001) & 0.043 & (0.001) & 0.013 & (0.001) \\ 
      
      \addlinespace[3pt]
      \\[-1.8ex]\hline
      \hline \\[-1.8ex]
    \end{tabular}
    \end{adjustbox}
    \caption{Average characteristics of units in each leaf, obtained by regressing each covariate on a set of dummies denoting leaf membership using only the honest sample. Standard errors are estimated via the Eicker-Huber-White estimator. Leaves are sorted in increasing order of the GATEs.}
    \label{table_average_characteristics_leaves_full}
    \end{table}
\endgroup 

\begingroup
  \setlength{\tabcolsep}{8pt}
  \renewcommand{\arraystretch}{1.2}
  \begin{table}[H]
    \centering
    \begin{adjustbox}{width = 1\textwidth}
    \begin{tabular}{@{\extracolsep{5pt}}l c c c c}
      \\[-1.8ex]\hline
      \hline \\[-1.8ex] 

      & \textit{Leaf 1} & \textit{Leaf 2} & \textit{Leaf 3} & \textit{Leaf 4} \\
      \addlinespace[2pt]
      \hline \\[-1.8ex] 

      \multirow{2}{*}{GATEs} & -252.941 & -205.349 & -182.874 & -176.181 \\
      & [-265.761, -240.121] & [-237.959, -172.739] & [-198.666, -167.082] & [-189.003, -163.359] \\

      \addlinespace[2pt]
      \hline \\[-1.8ex] 

      \textit{Leaf 1} & - & - & - & - \\
            & (-) & (-) & (-) & (-) \\ 
      \textit{Leaf 2} & 47.59 &    - &    - & - \\
            & (0.031) & (-) & (-) & (-) \\ 
      \textit{Leaf 3} & 70.07 & 22.48 &    - & - \\
            & (0.000) & (0.448) & (-) & (-) \\ 
      \textit{Leaf 4} & 76.76 & 29.17 &  6.69 & - \\
            & (0.000) & (0.308) & (0.519) & (-) \\ 

      \addlinespace[3pt]
      \\[-1.8ex]\hline
      \hline \\[-1.8ex]
    \end{tabular}
    \end{adjustbox}
    \caption{Point estimates and $95\%$ confidence intervals for the GATEs. Leaves are sorted in increasing order of the GATEs. Additionally, the GATE differences across all pairs of leaves are displayed. $p$-values testing the null hypothesis that a single difference is zero are adjusted using Holm's procedure and reported in parenthesis under each point estimate.}
    \label{table_differences_gates_four}
    \end{table}
\endgroup 

\begingroup
  \setlength{\tabcolsep}{8pt}
  \renewcommand{\arraystretch}{1.2}
  \begin{table}[b!]
    \centering
    \begin{adjustbox}{width = 1\textwidth}
    \begin{tabular}{@{\extracolsep{5pt}}l c c c c c c}
      \\[-1.8ex]\hline
      \hline \\[-1.8ex] 

      & \textit{Leaf 1} & \textit{Leaf 2} & \textit{Leaf 3} & \textit{Leaf 4} & \textit{Leaf 5} & \textit{Leaf 6} \\
      \addlinespace[2pt]
      \hline \\[-1.8ex] 

      \multirow{2}{*}{GATEs} & -269.937 & -242.638 & -205.349 & -193.974 & -176.181 & -151.200 \\
      & [-289.118, -250.756] & [-259.631, -225.645] & [-237.959, -172.739] & [-212.363, -175.585] & [-189.003, -163.359] & [-182.001, -120.399] \\

      \addlinespace[2pt]
      \hline \\[-1.8ex] 

      \textit{Leaf 1} & - & - & - & - & - & - \\
            & (-) & (-) & (-) & (-) & (-) & (-) \\ 
      \textit{Leaf 2} & 27.30 &    - &    - &    - &    - & - \\
            & (0.221) & (-) & (-) & (-) & (-) & (-) \\ 
      \textit{Leaf 3} & 64.59 & 37.29 &    - &    - &    - & - \\
            & (0.007) & (0.234) & (-) & (-) & (-) & (-) \\ 
      \textit{Leaf 4} & 75.96 & 48.66 & 11.38 &    - &    - & - \\
            & (0.000) & (0.001) & (0.551) & (-) & (-) & (-) \\ 
      \textit{Leaf 5} & 93.76 & 66.46 & 29.17 & 17.79 &    - & - \\
            & (0.000) & (0.000) & (0.411) & (0.411) & (-) & (-) \\ 
      \textit{Leaf 6} & 118.74 &  91.44 &  54.15 &  42.77 &  24.98 & - \\
            & (0.000) & (0.000) & (0.144) & (0.144) & (0.411) & (-) \\ 

      \addlinespace[3pt]
      \\[-1.8ex]\hline
      \hline \\[-1.8ex]
    \end{tabular}
    \end{adjustbox}
    \caption{Point estimates and $95\%$ confidence intervals for the GATEs. Leaves are sorted in increasing order of the GATEs. Additionally, the GATE differences across all pairs of leaves are displayed. $p$-values testing the null hypothesis that a single difference is zero are adjusted using Holm's procedure and reported in parenthesis under each point estimate.}
    \label{table_differences_gates_six}
    \end{table}
\endgroup 

\begin{figure}[H]
    \centering
    \includegraphics[scale=0.4]{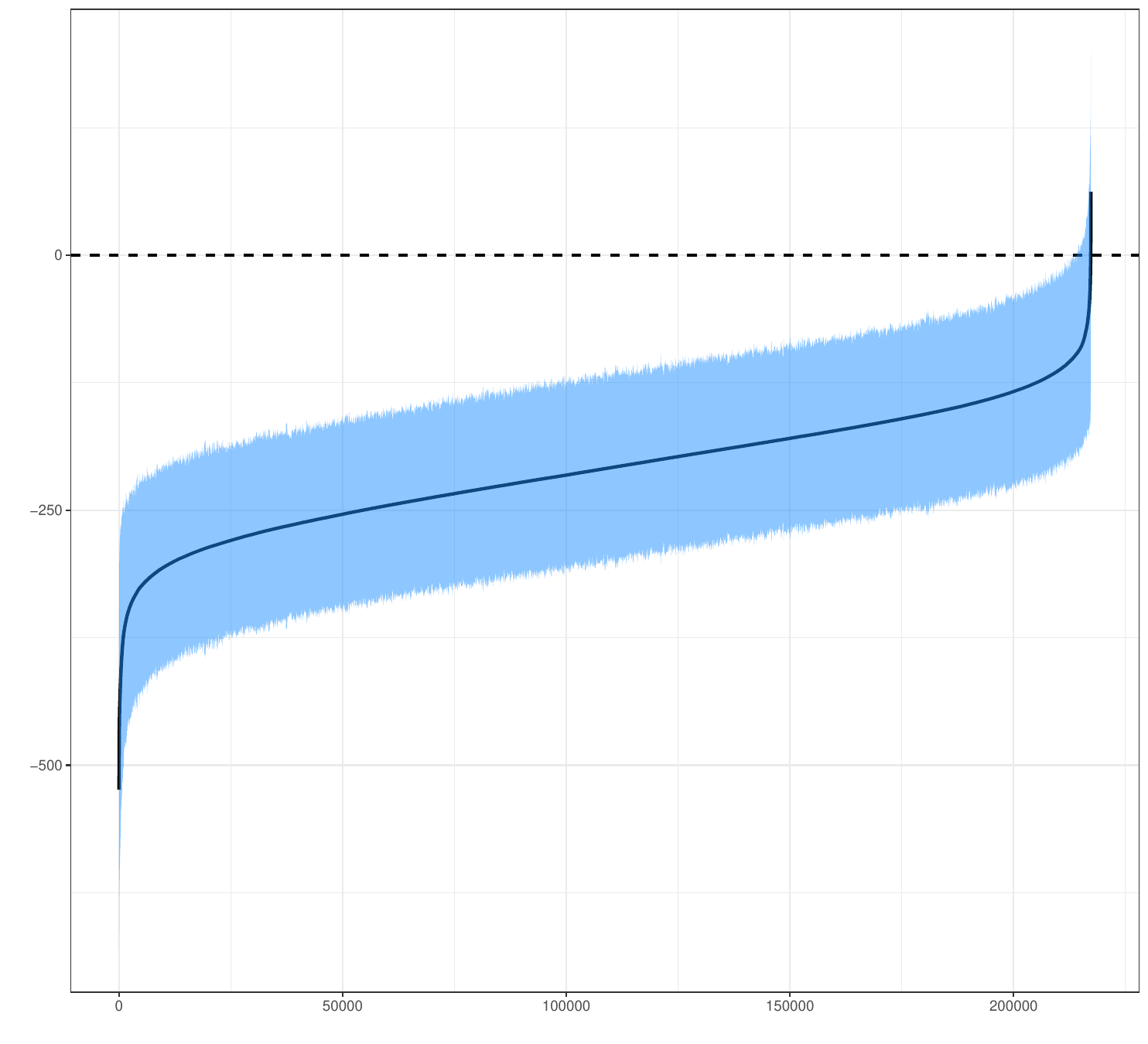}
    \caption{Sorted CATEs and 95\% confidence intervals. Predictions on the honest sample are shown. Standard errors are smoothed by a Nadaraya-Watson regression.}
    \label{fig_sorted_cates}
\end{figure}

\begin{figure}[H]
  \centering
  \begin{subfigure}[t]{0.48\textwidth}
    \centering
    \includegraphics[width=\linewidth]{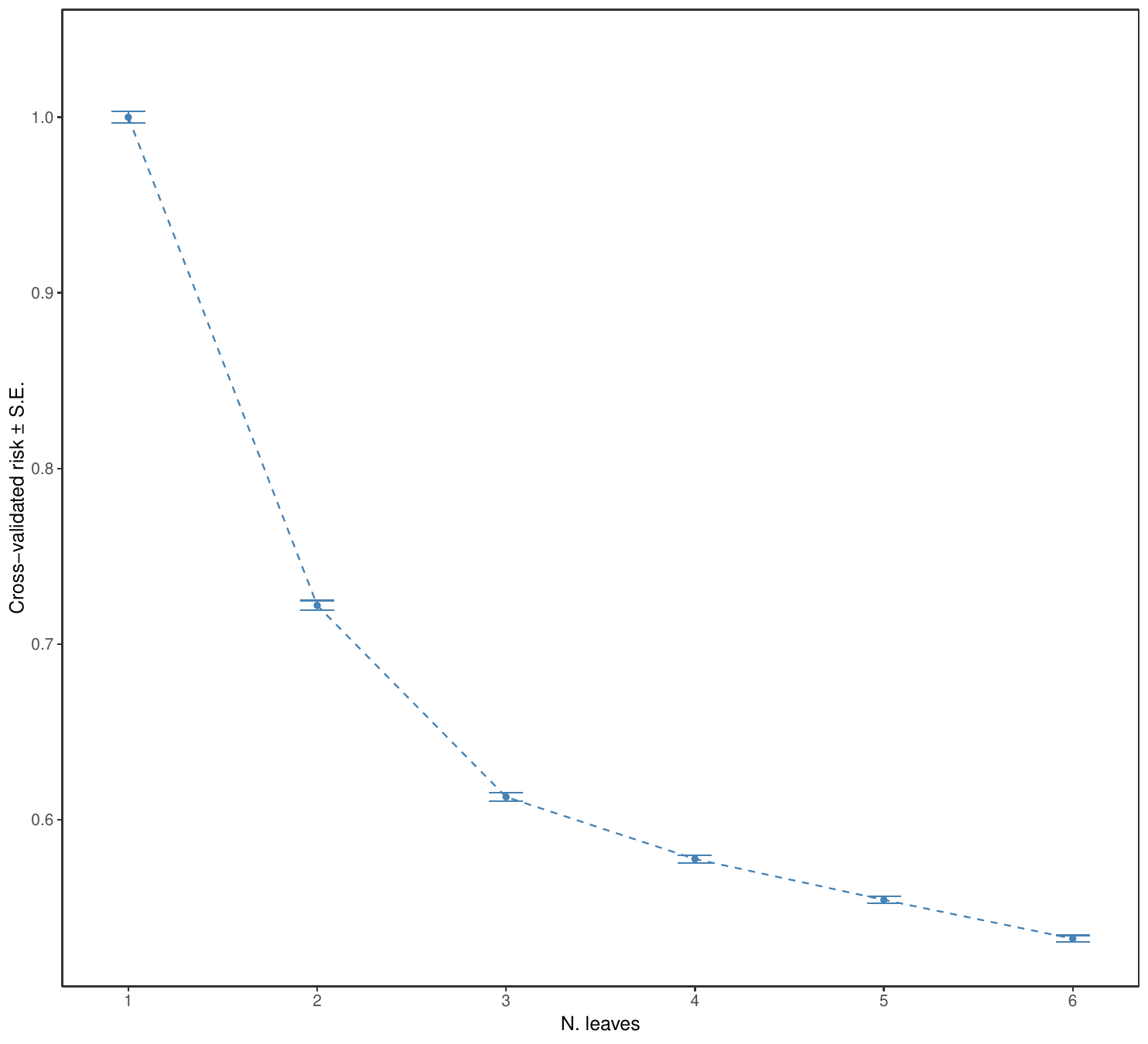}
  \end{subfigure}\hfill
  \begin{subfigure}[t]{0.48\textwidth}
    \centering
    \includegraphics[width=\linewidth]{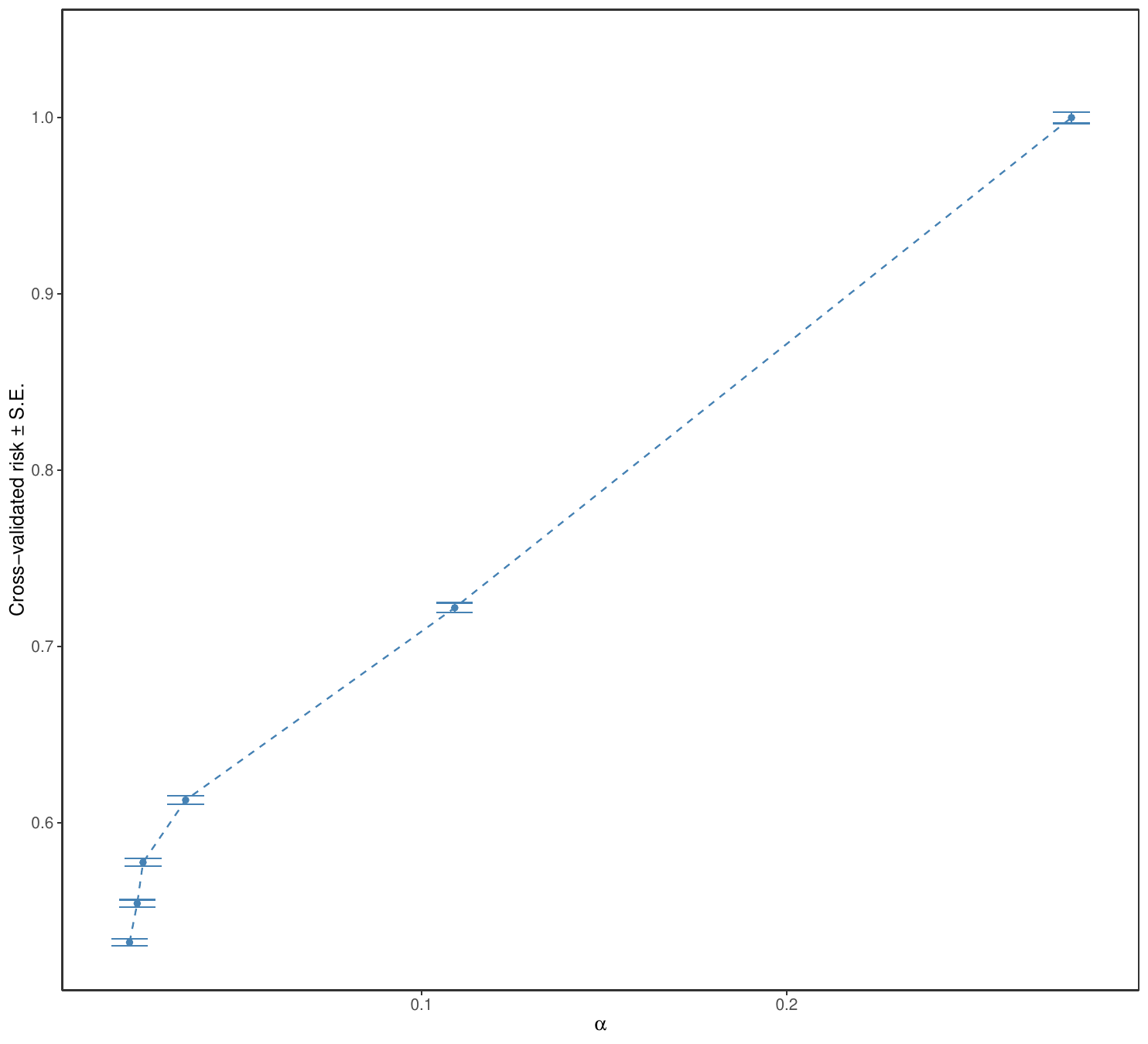}
  \end{subfigure}
  \caption{Cross-validated risk along the pruning path in Figure \ref{fig_sequence_groupings}. Points show mean cross-validated risk with $\pm$ S.E.\ bars. The left panel uses the number of leaves on the horizontal axis; the right panel uses the cost–complexity parameter $\alpha$.}
  \label{fig_alpha_path_dual}
\end{figure}

\newpage

%%% FURTHER SIMULATION RESULTS.
\section{Further simulation results}
\label{app_further_simulation_results}

\setcounter{equation}{0}
\setcounter{table}{0}
\setcounter{figure}{0}

\begingroup
  \setlength{\tabcolsep}{8pt}
  \renewcommand{\arraystretch}{1.1}
  \begin{table}[H]
     \centering
     \begin{adjustbox}{width = 1\textwidth}
     \begin{tabular}{@{\extracolsep{5pt}}l c c c c c c c c c c c c }
     \\[-1.8ex]\hline
     \hline \\[-1.8ex]
     & \multicolumn{6}{c}{$D_i \sim \textit{Bernoulli} \left( 0.5 \right)$} & \multicolumn{6}{c}{$D_i \sim \textit{Bernoulli} \left( \estimatedPropensityScore{X_i} \right)$} \\ \cmidrule{2-7} \cmidrule{8-13} 
     & \multicolumn{3}{c}{$a = 20$} & \multicolumn{3}{c}{$a = 50$} & \multicolumn{3}{c}{$a = 20$} & \multicolumn{3}{c}{$a = 50$} \\ \cmidrule{2-4} \cmidrule{5-7} \cmidrule{8-10} \cmidrule{11-13} 
     & 500 & 1,000 & 2,000 & 500 & 1,000 & 2,000 & 500 & 1,000 & 2,000 & 500 & 1,000 & 2,000 \\ 
     \addlinespace[2pt]
     \hline \\[-1.8ex] 

     \multicolumn{12}{l}{\textbf{\small Panel 1: $\overline{RMSE}$}} \\
     $AT_{\scriptstyle_{XL}}$ & 280.84 & 223.02 & 167.98 & 286.88 & 227.09 & 176.50 & 331.85 & 264.74 & 205.49 & 336.30 & 267.95 & 212.67 \\
     $AT_{\scriptstyle_{CF}}$ & 256.06 & 207.57 & 163.08 & 258.09 & 213.20 & 169.13 & 268.20 & 224.43 & 184.71 & 275.93 & 235.53 & 192.87 \\
     $CT$ & 137.92 & 94.53 & 57.36 & 163.90 & 111.22 & 75.63 & 341.11 & 334.26 & 325.07 & 429.13 & 450.82 & 469.35 \\ \cmidrule{1-13} 

     \multicolumn{12}{l}{\textbf{\small Panel 2: $\overline{\left\vert Bias \right\vert}$}} \\
     $AT_{\scriptstyle_{XL}}$ & 17.49 & 16.92 & 16.56 & 40.38 & 40.31 & 39.92 & 23.80 & 21.41 & 21.16 & 44.17 & 45.66 & 43.21 \\
     $AT_{\scriptstyle_{CF}}$ & 16.94 & 16.78 & 16.44 & 40.31 & 40.35 & 39.83 & 24.21 & 21.99 & 20.75 & 45.73 & 46.2 & 43.13 \\
     $CT$ & 16.39 & 16.14 & 15.96 & 40.15 & 39.95 & 39.75 & 49.96 & 45.32 & 43.40 & 84.52 & 78.65 & 87.62 \\ \cmidrule{1-13} 

     \multicolumn{12}{l}{\textbf{\small Panel 3: $\overline{SD}$}} \\
     $AT_{\scriptstyle_{XL}}$ & 279.96 & 221.97 & 166.63 & 282.30 & 221.28 & 169.08 & 330.51 & 263.38 & 203.79 & 331.57 & 261.73 & 205.61 \\
     $AT_{\scriptstyle_{CF}}$ & 255.17 & 206.49 & 161.75 & 253.09 & 207.12 & 161.63 & 266.46 & 222.71 & 182.83 & 269.84 & 228.29 & 185.15 \\
     $CT$ & 136.29 & 92.01 & 53.17 & 155.88 & 99.37 & 57.42 & 336.22 & 330.12 & 321.08 & 417.16 & 440.76 & 457.51 \\ \cmidrule{1-13} 

     \multicolumn{12}{l}{\textbf{\small Panel 4: Coverage of $95\%$ CI}} \\
     $AT_{\scriptstyle_{XL}}$ & 0.74 & 0.72 & 0.70 & 0.74 & 0.72 & 0.70 & 0.82 & 0.79 & 0.76 & 0.80 & 0.78 & 0.76 \\
     $AT_{\scriptstyle_{CF}}$ & 0.79 & 0.75 & 0.72 & 0.79 & 0.75 & 0.71 & 0.86 & 0.83 & 0.79 & 0.85 & 0.82 & 0.78 \\
     $CT$ & 0.69 & 0.71 & 0.76 & 0.66 & 0.68 & 0.75 & 0.60 & 0.58 & 0.58 & 0.61 & 0.61 & 0.62 \\ \cmidrule{1-13} 

     \multicolumn{12}{l}{\textbf{\small Panel 5: $\overline{|\mathcal{T}|}$}} \\
     $AT_{\scriptstyle_{XL}}$ & 11.79 & 13.59 & 14.32 & 12.00 & 13.47 & 14.47 & 10.40 & 12.02 & 13.37 & 10.38 & 11.84 & 13.30 \\
     $AT_{\scriptstyle_{CF}}$ & 11.20 & 12.87 & 14.02 & 11.10 & 12.95 & 13.91 & 8.87 & 10.70 & 12.52 & 8.91 & 10.80 & 12.37 \\
     $CT$ & 2.02 & 1.75 & 1.46 & 2.50 & 1.99 & 1.55 & 6.35 & 11.40 & 20.59 & 11.68 & 25.79 & 58.77 \\ 

     \hline
     \hline \\[-1.8ex]
  \end{tabular}
  \end{adjustbox}
  \caption{Comparison with causal trees. The first three panels report the average over the validation sample of $RMSE$ ($\overline{RMSE}$), $\left\vert Bias \right\vert$ ($\overline{\left\vert Bias \right\vert}$), and $SD$ ($\overline{SD}$). The fourth panel reports coverage rates for $95\%$ confidence intervals. The last panel reports the average number of leaves. All trees are adaptive.}
  \label{table_simulation_results_adaptive} 
  \end{table}
\endgroup

\begin{figure}[H]
    \centering
    \includegraphics[scale=0.5]{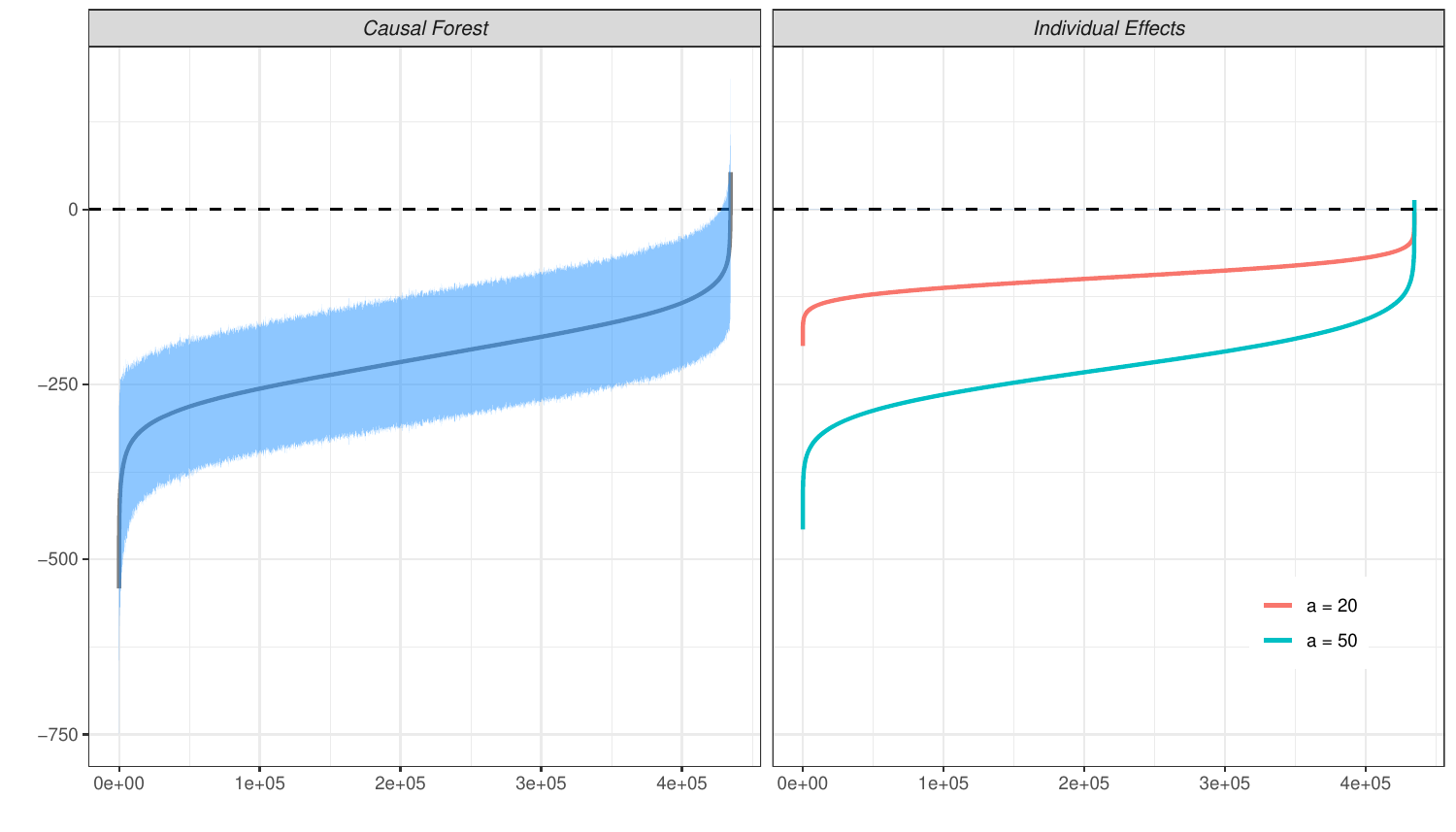}
    \caption{Sorted CATEs and 95\% confidence intervals (left panel) and individual effects (right panel). The CATEs are estimated via an honest causal forest, and standard errors are smoothed by a Nadaraya-Watson regression. The individual effects are constructed using model (\ref{equation_model_individual_effects}) with two distinct values of $a$.}
    \label{fig_ites_cates}
\end{figure}

\newpage

%%% BOUNDING THE NUMBER OF TREES.
\section{Bounding the Number of Trees}
\label{app_n_trees}

\setcounter{equation}{0}
\setcounter{table}{0}
\setcounter{figure}{0}

\begin{theorem}
    Define the \textit{\open depth"} $\mathcal{D}$ of a binary decision tree as the number of nodes connecting the root to the furthest leaf. Let $X \in \covariateSpace$ be a $p$-vector of binary covariates. Then, the number of distinct decision trees constructed by recursively partitioning $\covariateSpace$ and having a depth equal to or lower than $\mathcal{D}$ is bounded from below by $L_{\mathcal{D}} = \prod_{d = 1}^{\mathcal{D}} ( p - ( d - 1 ) )^{2^{d - 1}}$.
    \label{theorem_bound_trees}
\end{theorem}

\begin{proof}
    The proof is a matter of careful counting and relies on the fundamental theorem of counting. Define a \textit{symmetric $\mathcal{D}$-depth tree} as any binary decision tree such that the number of nodes connecting the root to each leaf equals $\mathcal{D}$. The root is considered a 0-depth tree. 
    
    Start from the whole covariate space $\covariateSpace$, i.e., from the unique 0-depth tree. Since all the $p$ covariates are binary, there is a unique candidate splitting point $s$ for each. Therefore, there exist $p$ distinct candidate pairs $( j, s )$ for the first split. It follows that it is possible to build $p$ distinct symmetric 1-depth trees.
    
    Now, fix a symmetric 1-depth tree, assuming without loss of generality that the split occurred on the first covariate. A symmetric 2-depth tree is then obtained by splitting both leaves of the nested symmetric 1-depth tree. As a split already occurred on the first covariate, there exist $p - 1$ distinct candidate pairs $( j, s )$ for splitting each terminal node. Therefore, from a given symmetric 1-depth tree it is possible to build $( p - 1 )^2$ distinct symmetric 2-depth trees. By the fundamental theorem of counting, the number of distinct symmetric 2-depth trees equals $p ( p - 1 )^2$.

    By a similar argument, it is easy to count the number of distinct symmetric 3-depth trees that can be constructed from any symmetric 2-depth tree, which equals $( p - 2 )^4$. Again, from the fundamental theorem of counting it follows that the number of distinct symmetric 3-depth trees equals $p ( p - 1 )^2 ( p - 2 )^4$.
    
    Iterating the argument, we can write a closed-form expression of the number of symmetric $\mathcal{D}$-depth trees that can be constructed using $p$ binary covariates:
    
    \begin{equation}
        L_{\mathcal{D}} = \prod_{d = 1}^\mathcal{D} ( p - ( d - 1 ) )^{2^{d - 1}}
        \label{equation_bound_trees}
    \end{equation}
    
    Notice that any binary decision tree with a depth equal to or lower than $\mathcal{D}$ can be regarded as a subtree of a given symmetric $\mathcal{D}$-depth tree, that is, it can be obtained by collapsing a certain number of internal nodes of the latter. Therefore, the set of symmetric $\mathcal{D}$-depth trees is a subset of all the possible distinct binary decision trees that can be constructed by recursively partitioning $\covariateSpace$ whose depth is at most $\mathcal{D}$. It follows that $L_{\mathcal{D}}$ is a lower bound for the number of such trees.
\end{proof}

\noindent \textit{Remarks.} Equation (\ref{equation_bound_trees}) has a nice interpretation. Notice that a symmetric $\mathcal{D}$-depth tree is composed of $2^{\mathcal{D}}$ terminal nodes. Therefore, the formula reflects the fact that starting from any symmetric $( d - 1 )$-depth tree, $2^{d - 1}$ leaves must be split to form a symmetric $d$-depth tree, and that $p - ( d - 1 )$ candidate pairs $( j, s )$ exist for each of these splits. 

Notice also that we cannot grow symmetric trees with depth $\mathcal{D} > p$: in such cases, $L_{\mathcal{D}} = 0$. Moreover, $L_{p - 1} = L_{p}$: starting from any symmetric $( p - 1 )$-depth tree, each leaf can be split choosing one and only one candidate pair $( j, s )$, hence only one symmetric $p$-depth tree can be constructed for each of the distinct symmetric $( p - 1 )$-depth trees.
    
In the case of $p$ categorical covariates with $k$ categories each, Theorem \ref{theorem_bound_trees} holds if we substitute $p ( k - 1 )$ for $p$. 

\end{appendices}

\end{document}